\documentclass[twocolumn]{aastex631}
\usepackage{amsmath,amstext}
\usepackage[T1]{fontenc}
\usepackage{CJKutf8}
\usepackage{comment}
\usepackage{hyperref}
\usepackage{enumitem}
\usepackage{booktabs}
\usepackage{caption}
\usepackage{graphicx}
\usepackage[normalem]{ulem}
\shortauthors{Collyer et al.}

\begin{document}

\title{The Surfaces of Small to Mid-Size Plutinos: Evidence of an Association Between Inclination and Surface Type}

\author[0009-0004-7149-5212]{Cameron Collyer}
\affiliation{Florida Space Institute, University of Central Florida, 12354 Research Parkway, Orlando, FL 32826, USA}
\affiliation{Department of Physics, University of Central Florida, 4111 Libra Drive, Orlando, FL 32816, USA}
\affiliation{Center for Astrophysics | Harvard \& Smithsonian; 60 Garden Street, Cambridge, MA, 02138, USA}

\author[0000-0003-4797-5262]{Rosemary~E. Pike}
\affiliation{Center for Astrophysics | Harvard \& Smithsonian; 60 Garden Street, Cambridge, MA, 02138, USA}

\author[0000-0001-7244-6069]{Ying-Tung Chen (\begin{CJK*}{UTF8}{bkai}陳英同\hspace{-0.15cm}\end{CJK*})}
\affiliation{Institute of Astronomy and Astrophysics, Academia Sinica; 11F of AS/NTU Astronomy-Mathematics Building, No. 1 Roosevelt Rd., Sec. 4, Taipei 10617, Taiwan}

\author[0000-0003-4143-8589]{Mike Alexandersen}
\affiliation{Center for Astrophysics | Harvard \& Smithsonian; 60 Garden Street, Cambridge, MA, 02138, USA}

\author[0009-0007-9017-4010]{Mark~Comte}
\affiliation{Campion College and the Department of Physics, University of Regina, Regina, SK S4S 0A2, Canada}

\author[0000-0001-5368-386X]{Samantha~M. Lawler}
\affiliation{Campion College and the Department of Physics, University of Regina, Regina, SK S4S 0A2, Canada}

\author[0000-0002-6117-0164]{Bryan~J. Holler}
\affiliation{Space Telescope Science Institute, 3700 San Martin Drive, Baltimore, MD 21218, USA}

\author[0000-0001-7032-5255]{J.~J. Kavelaars}
\affiliation{Herzberg Astronomy and Astrophysics Research Centre, National Research Council of Canada, 5071 West Saanich Rd, Victoria, BC V9E 2E7, Canada}
\affiliation{Department of Physics and Astronomy, University of Victoria, Victoria, BC, Canada}

\author[0000-0002-9179-8323]{Lowell Peltier}
\affiliation{Department of Physics and Astronomy, University of Victoria, Victoria, BC, Canada}

\begin{abstract} 
\noindent Being one of the most populated mean motion resonances (MMR) with Neptune and lying close to the inner boundary of the present day cold classical disk, observations of the orbital and surface class distributions of the plutinos in the 3:2 MMR provide constraints on Neptune's migration and insight into the compositional structure of the pre-migration planetesimal disk. Here, we present observations of the surface reflectance of 43 small to mid-size $(H_V \gtrsim 5)$ transneptunian objects (TNOs) through the $grz$ wavelength range, 14 of which are plutinos. We classify the surfaces of these TNOs using the two-surface class model (FaintIR and BrightIR surface classes) proposed by Fraser and collaborators, where the FaintIR surface class is dominated by cold classicals. Incorporating similar observations of plutinos from the literature for a total sample size of 43 plutinos, we find that the osculating inclination distributions ($i_{osc}$) of the FaintIR and BrightIR plutinos are statistically distinguishable at 99.3\% significance, where (6/7) of the plutinos with $i_{osc} < 4.5$° have FaintIR surfaces. This is most easily explained if the FaintIR and BrightIR planetesimals were radially partitioned in the primordial planetesimal disk before being captured into the 3:2 resonance. While this could be evidence that the primordial cold classical disk with FaintIR surfaces was broader in the past by $\sim$3 au in the sunward direction, we cannot rule out the alternative explanation that these FaintIR plutinos were scattered from the $\sim 42.5 - 48$ au region from the Sun and captured into the 3:2 resonance.
\end{abstract}

\section{Introduction}
In the current framework, the present-day transneptunian population is composed of a suite of planetesimals that formed beyond $30$ au, permeated by a collection of planetesimals that were gravitationally emplaced onto their current orbits from the region $\sim 24-30$ au from the Sun during Neptune's migration \citep{gomes2003, levison2008, morbidelli2008, Gladman2021}. The physical characteristics of the pre-migration transneptunian planetesimal disk, such as its surface mass density profile and compositional stratification, as well as the mode of Neptune's migration, are widely debated topics in the community \citep[e.g.][]{malhotra1995, brown2011, nesvorny2015a,nesvorny2015b,nesvorny16, volkmalhotra2019, nesvorny2020, alidib2021, buchanan2022, Marsset23}.

Various lines of evidence suggest that the cold classical\footnote{See \cite{Gladman2021} for a recent review of TNO dynamical classifications.} planetesimal disk currently between $\simeq 42.5-48$ au was broader in heliocentric distance in the solar system's past. The current cold classical population is sharply bounded on its interior by the overlapping $\nu_8$ and $\nu_{18}$ secular resonances, currently located at $40-42$ au, that make this region extremely unstable \citep[e.g.][]{duncan95, morbidelli1995, morbidelli1997, lykawka2005}. During Neptune's migration, these secular resonances moved in step with Neptune, likely exciting and destabilizing many planetesimals in their path \citep[e.g.][]{gomes2000,nesvorny2015a}, carving the sharp boundary near $42$ au when migration ceased. This time period is also when the majority of the planetesimals currently in Neptune's mean motion resonances were captured \citep[e.g.][]{malhotra1995, lykawka2005, morbidelli2008, Nesvorny2018, volkmalhotra2019}. The primordial cold classicals have a number of unique physical and orbital properties\footnote{We refer to `primordial' cold classicals as the population of planetesimals that originally formed beyond $\sim 30$ au with the physical and orbital characteristics of the current cold classical population.}. As noted in \cite{Napier2022}, the cold classical $H_V$-mag distribution appears to not extend below $H_V \simeq 5.65$ mag. Their most telling identifier for observers has been their optically red surfaces \citep{tegler2000,trujillobrown2002} and unique binary characteristics \citep{noll2008, Noll2018}. With knowledge of their surface and binary characteristics, TNOs currently in dynamically excited dynamical classes can be identified as members of the primordial cold classical population \citep[e.g.][]{ty430, Sheppard2012, tegler2016}. 

Because of their vast geocentric distances, surface studies of $H_V \gtrsim 5$ TNOs have been conducted in large through photometric colors or spectrophotometry spanning the optical \citep[e.g.][]{luu96,tegler1998,Sheppard2012} to the nearest infrared \citep[e.g.][]{Hainaut02, Hainaut_2012, pike2017, schwamb2019, seccull2021, Fraser2surfaces}. The reflectance of small to mid-size TNOs in the range of $\sim 0.35$ to $1.25$ $\mu m$ is well approximated by an optical spectral slope spanning $\sim 0.35$ to $0.8$ $\mu m$ and an optical near-IR spectral slope from $\sim 0.8$ to $1.25$ $\mu m$. 
Photometric colors sampling these two regions are therefore adequate for estimating their surface spectral reflectance in this wavelength range. Various surface taxonomy systems for TNOs have been proposed using surface colors and geometric albedos \citep[e.g.][]{fulchignoni2008, dalleore2013, lacerda14, belskaya2015, pike2017, Fraser2surfaces}. 
The simple two-surface class system as introduced by \cite{Fraser2surfaces} well models the bifurcated ($g-r$) \& ($r-J$) surface colors of the TNOs from the Colours of the Outer Solar System Origins Survey \citep[Col-OSSOS][]{schwamb2019}. 
In the \citet{Fraser2surfaces} surface classification scheme, the `FaintIR' surface class is dominated by the cold classical TNOs, providing a compelling framework for further observations.

The gravitational dispersal of the planetesimals from the original transneptunian population is largely thought to have been a stochastic process. However, substantial direct observations of dynamically excited TNOs with cold classical characteristics, located nearby the current cold classical region, can be significant evidence that the cold classical planetesimal disk was further extended in the past. The plutinos\footnote{Named after the largest member, Pluto.} in the 3:2 mean motion resonance are perhaps the most populated resonance with Neptune \citep{volk2016}. Located at $\sim39.4$ au, the plutinos are currently separated from the present day cold classical population by the dynamically unstable region at $40-42$ au. In the context of Neptune's migration, the 3:2 mean motion resonance could have captured primordial cold classicals near $39$ au that survived the sweeping $\nu_8$ and $\nu_{18}$ secular resonances \citep[e.g.][]{nesvorny2015a}. 

In this work, we present observations and surface classifications of 43 small to mid-size TNOs ($H_V \gtrsim 5$) whose reflectance in the optical Near-IR are measured for the first time, 14 of which are plutinos. Using the \citet{Fraser2surfaces} two-surface classification scheme, we incorporate optical Near-IR observations of plutinos from the literature, resulting in a collective sample size of 43 plutinos. We then compare the physical and orbital properties of the plutinos in each surface class and find an association between surface class and osculating inclination. Specifically, in Section \ref{sampledclass} we outline our target sample of plutinos as well as the main $grz$ sample, where we incorporate the Col-OSSOS $grz$ sub-sample. We then detail the adopted dynamical classification scheme. In Section \ref{sec:observations} the observations, reductions, and photometry technique are presented, followed by Section \ref{measuringcolors} where we describe the techniques used to quantify surface reflectance in the wavelength range considered. Section \ref{sec:dataproducts} presents the results from the $grz$ sample, and ties in observations of plutino surfaces from the literature in the context of the FaintIR/BrightIR surface classification scheme introduced in \citet{Fraser2surfaces}. Evidence for the association between inclination and surface type in the small to mid-size plutino population is detailed in Section \ref{main result}. Finally, we discuss the nature of this association and what it reveals about the compositional structure of the pre-migration planetesimal disk in Section \ref{conclusions}.

\section{Sample and Dynamical Classifications}
\label{sampledclass}

\subsection{Sample Description}
\label{sample}
Our observing program was designed with the goal of observing a brightness-complete ($m_r \leq 24.0$) sub-sample of plutinos from the Canada France Ecliptic Plane Survey \cite[CFEPS,][]{petit11} and the Large inclination Distant Objects survey \cite[L\textit{i}DO,][]{Mike_LiDO} as well as additional TNOs as comparison targets. Both the CFEPS and the L\textit{i}DO surveys used the Canada France Hawaii Telescope (CFHT) and the MegaPrime/MegaCam imager \citep{MegaCam_CFHT} for TNO discoveries. CFEPS primarily operated through $g$\footnote{Filter bandpass labels ($g$, $r$, $i$, $z$) denote the Sloan Digital Sky Survey Photometric System \citep{SDSS_photometric_system_old}.}-band, while the L\textit{i}DO survey used a wide band $gri$ filter.
 
To determine our $m_r \leq $ 24.0 target sample of plutinos from these two surveys, we assumed an a-priori ($g-r$) color of 0.6 mags. The CFEPS plutinos were converted to $r$-band by $r = g - (g-r)$. The L\textit{i}DO plutinos were converted to $r$-band by the color-transformation equations published in \citet{TonryPS1}, assuming the filter/detector response with the Pan-STARRS 1 $w_{PS1}$ filter is similar to that of the Canada France Hawaii Telescope MegaCam $w_{gri}$ filter. Table \ref{ttable} shows the plutinos from these two surveys with $m_r \leq 24.0$. All 24 plutinos from the CFEPS and six (of 21) plutinos from the L$i$DO survey are brighter than this limit. We specifically refer to these plutinos as the `target sample'. The brightest L$i$DO plutino excluded from our sample is $m_r$=24.11, $m_w=24.17$. In Table \ref{ttable}, we also present the status of our observing program. (14/30) plutinos from our target sample have been observed. Plutinos 2000 GN$_{171}$, 2001 KQ$_{77}$, 2004 KC$_{19}$, 2004 HX$_{78}$, 2004 VZ$_{130}$ and 2004 KB$_{19}$ are currently transiting the galactic plane, making accurate multi-band photometry nearly impossible due to nearby contaminating sources. To summarize, our magnitude-limited selection criteria from CFEPS and L$i$DO resulted in a target sample of 30 plutinos.

For the non-plutino comparison targets, a primary criteria in target selection for these objects was that they were visible during our classically scheduled observing runs and would provide sufficient SNR color measurements in $\sim 30$ minutes of observing, which was the observing sequence utilized for the brighter plutinos in our sample. We prioritized observing dynamically classified cold classical TNOs as comparison targets. This was done in order to constrain the full surface characteristics of cold classical TNOs in the $grz$ wavelength range and to guide the surface classification scheme in $grz$. A full description of the TNOs observed in this program is tabulated in Table \ref{tableA}.


In order to boost the sample size, we immediately incorporated into our main sample the Col-OSSOS sub-sample of TNOs from whose ($g-r$) \& ($r-z$) color photometry was published in \cite{pike2017z} and \cite{Fraser2surfaces}. Because of the random nature of the $z$-band Col-OSSOS sub-sample, the Col-OSSOS did not measure ($r-z$) colors of any plutinos, but did measure several cold classicals and dynamically excited TNOs. For simplicity we refer to the TNOs whose color photometry was measured by the present authors, as well as the Col-OSSOS  $z$-band sub-sample, collectively as the $grz$ sample. The $grz$ sample consists of 69 TNOs; 43 TNOs whose $grz$ photometry is reported here for the first time of which 14 are plutinos, and 26 non-plutino TNOs from the Col-OSSOS $z$-band sub-sample. 
 
A main objective of this study is to understand the full range and distribution of surface types within the small to mid-size plutino population. (341520) Mors–Somnus (Mors–Somnus hereafter) is a binary plutino whose components, Mors and Somnus, are widely separated and of similar size which is a common property of binary systems in the cold classical population \citep{Noll2018}. Moreover, initial $gri$ photometry revealed the surfaces of both components were optically red, resembling as well the surface colors of cold classical TNOs \citep{ty430}. Since $z$-band photometry pushes further into the Near-IR and is more informative of surface type, we observed Mors–Somnus in $grz$ to compare it to the other cold classicals and plutinos in the $grz$ sample. We also resolved Mors and Somnus separately and compared their reflectance through $grz$, which offers insight into the likely formation mechanism of the system \citep{Benecchi2009, marsset2020}. Recently, Mors and Somnus were observed to have indistinguishable surface compositions from $0.6$ to $5.3$ $\mu m$ spectra using the PRISM/CLEAR disperser on the JWST. They also show spectral similarity to cold classicals in this wavelength range \citep{carol2024}. Our $grz$ photometry of Mors-Somnus can therefore also serve as a link between $grz$ photometry and true surface composition. 

\subsection{Dynamical Classifications}
\label{dclass}
The TNOs considered in this work were dynamically classified following the \citet{gladman2008} and \citet{Gladman2021} classification scheme. This classification scheme is based on 10 Myr numerical simulations of each TNO to track their orbital evolution under the gravitational influence of the Sun and the giant planets, adding the mass of the terrestrial planets to the Sun. Orbits for the TNOs from the CFEPS, Col-OSSOS, and H/WTSOSS surveys/datasets were taken from \citet{volk_vanlaerhoven2024}, where for each TNO, three clones were simulated using the best-fit orbit and the minimum and maximum semi-major axis variation of its orbit. Orbits for the L$i$DO TNOs were determined by simulating 201 clones with re-sampled orbital parameters that reflect the range of uncertainty of the orbit-fits \citep{Mike_LiDO}. We adopted the dynamical classifications from \citet{volk_vanlaerhoven2024} and \citet{Mike_LiDO} exactly for TNOs classified to be in the following groups: Jupiter-coupled, centaur, resonant, scattering, detached. Free inclinations published in \citet{vanLaerhoven2019} and \citet{huang2022} were used to separate the classical TNOs as follows. Cold classical TNOs are defined to have $i_{free} < 4 ^{\circ}$ between $42.5 < a < 45$ au, and $i_{free} < 6 ^{\circ}$ between $45 < a < 47$ au, while hot classical TNOs have larger free inclinations in these two regions \citep{vanLaerhoven2019}. Dynamical classifications for the TNOs whose $grz$ photometry is presented here are tabulated in Table \ref{tableA}.

In this work, we group the Jupiter-coupled, centaur, scattering, detached and hot classical TNOs as the `non-resonant dynamically excited' population. TNOs in resonance with Neptune within the $42.5-47$ au region are collectively referred to as `resonant in the main belt'. The resonant TNOs with semi-major axes beyond $49$ au are collectively referred to as `resonant $a$ > 49 au'. 

After sorting by dynamical class, the main $grz$ sample is composed of (14) plutinos, (21) cold classicals, (19) non-resonant dynamically excited TNOs, (4) 4:3s, (7) resonant TNOs within the main belt, and (3) resonant TNOs with $a>49$ au. 

\section{Observations, Reductions and Photometry Technique}
\label{sec:observations}
The 6.5 m Magellan Baade Telescope located at Las Campanas Observatory in the Atacama Desert of Chile was used for all photometric observations. Observations reported here were gathered from 2020 December 22 through 2024 April 6. We primarily used chip 8 of the $f/4.3$ Mosaic2 E2V CCD camera on the Inamori-Magellan Areal Camera and Spectrograph \citep[IMACS,][]{IMACS2011}. The pixel scale of chip 8 is $0\overset{''}{.}110$px$^{-1}$ and $0\overset{''}{.}221$px$^{-1}$ when binned $1\times1$ and $2\times2$ respectively, with a field of view of $3\overset{'}{.}87\times7\overset{'}{.}63$ for the $2112\times4160$ chip. Due to electronics issues with the $f/4.3$ camera, we used the IMACS $f/2.5$ Mosaic3 E2V CCD camera for observations on the night of 2024 February 8, which has a pixel scale of 0$\overset{''}{.}200$ px$^{-1}$ when using $1\times1$ binning and a FOV of $7\overset{'}{.}04\times13\overset{'}{.}9$ for the $2112\times4160$ chip. The two cameras have identical layouts and a single filter set which is installed on the camera being used. The only significant difference is the focal length.

Sloan $g$, $r$ and $z$-band filters available for the IMACS instrument\footnote{\url{https://www.lco.cl/technical-documentation/imacs-user-manual/} contains filter specifications and throughputs.} were used in sequence for each visit. All images were reduced with routine flat and bias corrections using master flat and master bias frames created for each day of observations. Master flat frames and the master bias frame were created by median combining their respective image stacks. In comparing dome flat corrected images with twilight flat corrected images, we found that the twilight flats were more effective at removing the flat-field effects on the transmitted light at the 0.5-2\% level, band dependent. Accordingly, we only used dome flats for reductions in the rare cases where twilight flats were unavailable.

Images taken in the Near-IR with the $f/4.3$ Mosaic2 and $f/2.5$ Mosaic3 E2V CCD cameras show fringe patterns that can be corrected with master fringe frames that capture the structure of the fringe pattern. Master fringe frames were created on a night-by-night basis from dithered $z$-band science images of equal exposure time. The master fringe frames were scaled to the background of each $z$-band image and then subtracted. Individual $z$-band exposures were capped at 200s to limit the intensity of the fringing and total background counts.

We tracked at the sidereal rate to preserve the quality of the stellar point spread functions (PSFs), allowing the TNO to trail slightly. We most often used the imaging sequence of $r-g-zzz-g-r$, where the three $z$-band images were stacked in processing. For the fainter objects, we used an imaging sequence of $r-zz-gg-r-zzz-r$, where consecutive $z$ and $g$ images were stacked. Exposure times were determined using the \texttt{Las Campanas Observatory Exposure Time Calculator}\footnote{\url{https://www.lco.cl/lcoastronomers/~iss/lcoetc/html/lcoetc_sspec.html}} and tuned to an anticipated SNR of 30 in $g$ and $r$ bands, and $20-30$ in $z$-band depending on target magnitude.

Aperture photometry was measured using the Trailed Image Photometry in Python package \cite[TRIPPy,][]{trippy}. TRIPPy uses pill-shaped apertures for aperture photometry to account for the object's rate of motion when tracking sidereally.  We used the common method of using small apertures that avoided the wings of the TNO PSF and added aperture corrections measured from high SNR in-frame stars. In-frame calibration stars were chosen by human inspection to ensure removal of galaxies, stars with cosmic ray contamination, stars with overlapping PSFs, stars near filter imperfections, etc. 
These stars were used for both the PSF generation and for the photometric zero point calculation.

The Pan-STARRS1 \cite[PS1,][]{PS1} database was used to calibrate our photometry. We calculated linear filter transformation equations between PS1 $grz$ and IMACS $grz$ magnitudes, and SDSS $grz$ and IMACS $grz$ magnitudes using the methodology of \citet{schwamb2019}, as described in Appendix \ref{color terms}. Image zeropoints were calculated in the IMACS $grz$ filter system after converting stellar magnitudes from the PS1 system to IMACS instrumental magnitudes. 

\section{Calculating Colors, Spectral Slopes and Reddening Line Projections}
\label{measuringcolors}

Most TNOs and asteroids exhibit sinusoidal fluctuations in their reflected-lightcurves over the course of a rotational period. The sinusoidal shape in the lightcurves is primarily due to the nonspherical shapes of these small bodies. In transneptunian space, measured rotation periods range from 15.8 days \cite[Eris,][]{Bernstein2023} to 3.9 hr \cite[Haumea,][]{Rabinowitz2006}. For the smaller TNOs ($H \gtrsim$ 5), the average observed rotation period falls at $\sim$10 hr, while their peak to peak magnitude fluctuations are $\sim$0.3 mag \citep{Benecchi2013, Thirouin2018, MikeArotate2019, Ashton2023}. Non-simultaneous color measurements of small bodies must therefore account for rotation during an imaging sequence for accurate measurements.

We adopt the color measurement technique described in \citet{schwamb2019} and elaborate on it here. Objects are imaged in a multi-band sequence that are anchored by $r$-band images. As noted above, we used imaging sequences $r-g-zzz-g-r$ and $r-zz-gg-r-zzz-r$ for the brighter and fainter objects corresponding to $\sim 30$ min and $\sim 44$ min of elapsed time when accounting for overheads, or $\sim5$\% and $\sim7.3$\% of 10 hours, the average rotation period noted above, respectively.

We assume that rotation-induced lightcurve variations during an imaging sequence are accurately modeled by a straight line on these short timescales. We also assume that during this small amount of rotation, the TNO's ($g-r$) and ($r-z$) colors remain constant. A least-squares methodology was used to determine the ($g-r$) and ($r-z$) colors which produce the least scatter in the linear lightcurve fit. This methodology produced linear fits with low least-squares values and no apparent systematic deviations from linear. Using the transformation equations shown in Appendix \ref{color terms}, the IMACS ($g-r$) and ($r-z$) colors were converted to ($g-r$) and ($r-z$) colors in the SDSS photometric system \citep{SDSS_photometric_system_old}.

The ($g-r$) \& ($r-z$) colors were converted to $\mathcal{S}_{g \to r}$ and $\mathcal{S}_{r \to z}$ spectral slopes to produce spectral slopes plots. We use spectral slopes plots here as opposed to color-color plots because they enable a clearer picture of the underlying reflectance spectra. The conversion between color to spectral slope is as follows. First, we convert ($g-r$) and ($r-z$) colors to a relative reflectance spectrum. The reflectance from a TNO surface in a particular band is the ratio of incident ($F_{\odot, \lambda}$) to reflected light ($F_{\text{TNO}, \lambda}$): $R_{\lambda} = F_{\text{TNO}, \lambda} / F_{\odot, \lambda}$. Relative reflectance is the reflectance in one filter relative to another. We computed relative reflectance with respect to $r$, where $\mathcal{R}(\lambda) = R_{\lambda} / R_{r}$, and $\mathcal{R}(\lambda_r)=1$. In terms of colors
\begin{equation}
    \mathcal{R}_{r}(\lambda) = 10^{ 0.4[ (m_{\lambda} - m_r)_{\odot} - (m_{\lambda} - m_{r})_{\text{TNO}}] }.
\end{equation}
For $\mathcal{R}(\lambda_{g})$, we use $(m_{g} - m_r)_{\odot} = 0.44$ mags, and for $\mathcal{R}(\lambda_{z})$, we use $(m_{z} - m_r)_{\odot} = -0.14$ mags. The effective wavelengths for $\lambda_{g}$, $\lambda_{r}$ and $\lambda_{z}$ are from \citet{SDSS_effectivewavelengths}, and the solar colors are from SDSS DR12\footnote{\url{https://live-sdss4org-dr12.pantheonsite.io/algorithms/ugrizvegasun/}} \citep{Alam2015}. 
The spectral slope is defined as the percent reddening per 100 nm \citep{luu96}. The general form of the equation for the spectral slope from wavelengths $\lambda_1 \to \lambda_2$, where $\lambda_1 < \lambda_2$, is 
\begin{equation}
\mathcal{S}_{1 \to 2} = 100 \times \frac{\mathcal{R}_{r}(\lambda_2)-\mathcal{R}_{r}(\lambda_1)}{(\lambda_2 - \lambda_1 \hspace{0.06cm} \text{[nm]}) / 100 \hspace{0.06cm} \text{nm}} 
\end{equation}
and is dimensionless.

\begin{figure*}
\centering
 	\includegraphics[scale=0.6]{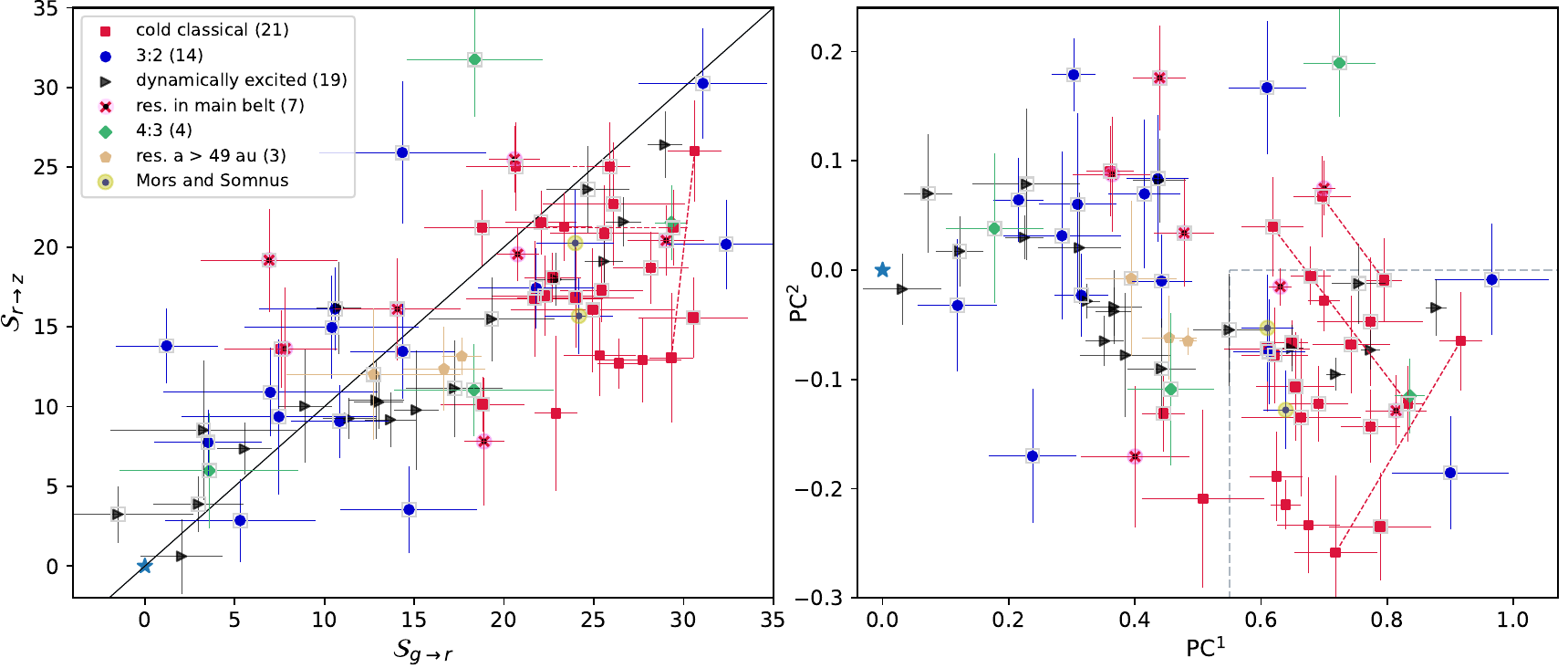}
	\caption{
    \textbf{Left:} $\mathcal{S}_{g \to r}$ \& $\mathcal{S}_{r \to z}$ spectral slopes for the TNOs in the $grz$ sample. The solid diagonal black line represents increasing spectral slopes corresponding to linear spectral reflectance through $grz$. \textbf{Right:} PC$^1$ \& PC$^2$ projections as described in Section \ref{measuringcolors}. The dashed gray lines define the two-dimensional range in which the FaintIR surface class is identified (see Section \ref{compclasses}). In both panels, the TNOs observed in our Magellan Baade program are outlined with grey squares. The other TNOs are the Col-OSSOS $grz$ sub-sample. In both panels, points connected by dashed red lines correspond to measurements of the same TNO at different epochs. Somnus is located above Mors in both panels.
 \label{figure:specslopes}
 }
\end{figure*}

Finally, we utilize the coordinate system proposed in \citet{Fraser2surfaces} for classifying surfaces of the TNOs in the $grz$ sample into FaintIR/BrightIR surface classes (see Section \ref{compclasses} for description of the FaintIR/BrightIR surface classification scheme). This new coordinate system is conceptually similar to a principle component analysis, where the two coordinates are defined by the projected distance along and away from the reddening line in color-color space\footnote{The software used for this coordinate conversion is available at  \url{https://github.com/fraserw/projection} \citep{Fraser2surfaces}.}. The use of the reddening line is motivated by the trend for icy objects to fall along this line in optical near-IR color-color space \cite[e.g.][]{Hainaut02, Hainaut_2012, Sheppard2012, peixinho2015, Fraser2surfaces}. The reddening line represents colors (or spectral slopes) of synthetic surfaces whose reflectance spectra are linear through the wavelengths of the filter set considered. A visual representation of the PC$^1$ and PC$^2$ axes in color-color space is shown in Figure C1 of \citet{Fraser2surfaces}. In our spectral slopes plots, the reddening line is the black diagonal line (e.g. Figure \ref{figure:specslopes} (left)). The re-projection of the the $\mathcal{S}_{g \to r}$ \& $\mathcal{S}_{r \to z}$ spectral slopes into PC$^1$ \& PC$^2$ aids in simplifying the FaintIR/BrightIR surface classification scheme in $grz$ and is shown in Figure \ref{figure:specslopes} (see Section \ref{compclasses} for further details on surface classifications).

\begin{figure*}
\centering
 	\includegraphics[scale=0.6]{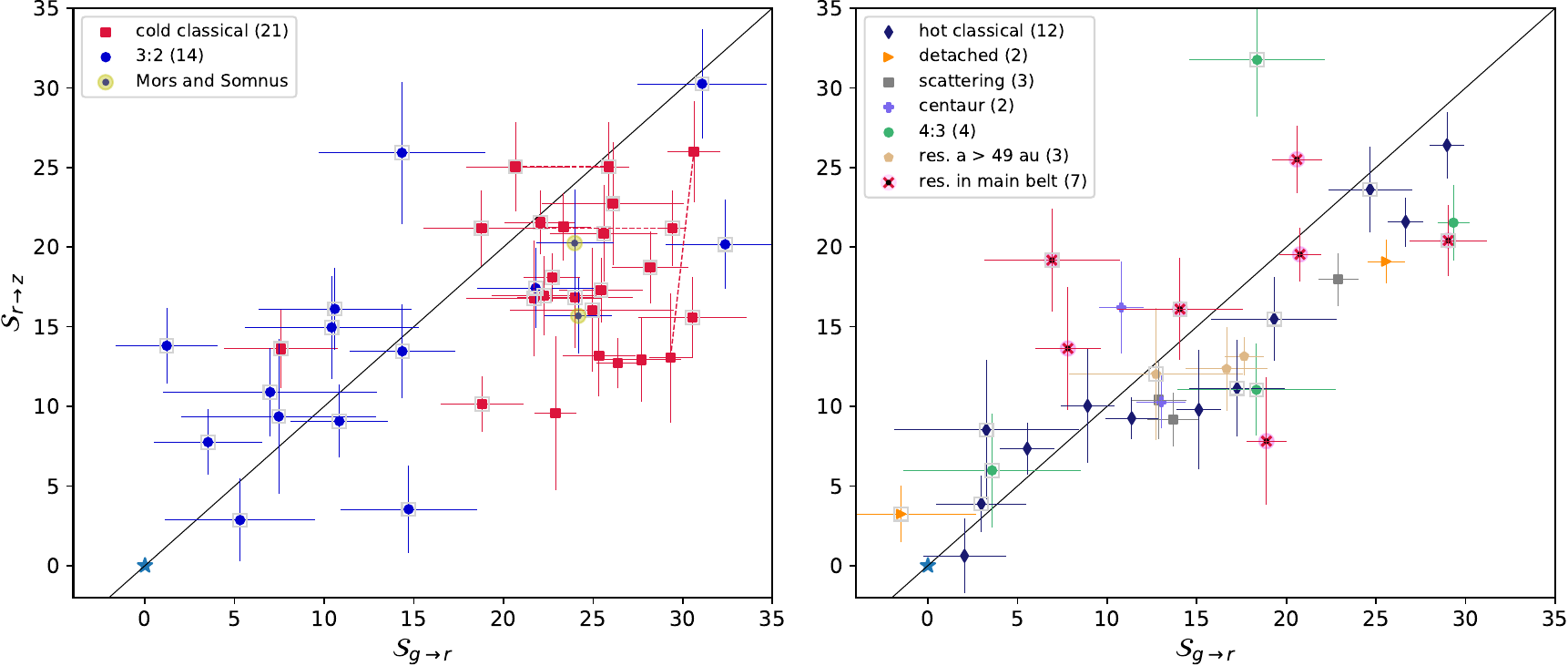}
	\caption{
 $\mathcal{S}_{g \to r}$ \& $\mathcal{S}_{r \to z}$ spectral slopes shown in Figure \ref{figure:specslopes} (left) broken down by dynamical class as noted in the legend of each panel. In both panels, the TNOs observed in our Magellan Baade program are outlined with grey squares. Points connected by dashed red lines correspond to measurements of the same TNO at different epochs. Somnus is located above Mors in the left panel.
 \label{figure:specslopes_bygroup}
 }
\end{figure*}

\section{Results}
\label{sec:dataproducts}

In this section we outline $\mathcal{S}_{g \to r}$ and $\mathcal{S}_{r \to z}$ spectral slope measurements, PC values and ($g-z$) colors for the TNOs in the $grz$ sample. In the following sections we provide context and the basis to interpret these results. In particular, see Section \ref{compclasses} for a description of the FaintIR/BrightIR classification scheme. 

In Figure \ref{figure:specslopes} (left) we present $\mathcal{S}_{g \to r}$ and $\mathcal{S}_{r \to z}$ spectral slopes of 69 TNOs; 43 are from our Magellan Baade program and 26 are from the Col-OSSOS. In Figure \ref{figure:specslopes} (right), the data is converted to components along a different set of orthogonal coordinates, PC$^1$ and PC$^2$, as described in Section \ref{measuringcolors}. For ease of interpretation, we separated the data in Figure \ref{figure:specslopes} (left) by dynamical class shown in Figure \ref{figure:specslopes_bygroup}, where the plutinos and cold classicals are on the left and TNOs in other dynamical classes are on the right. The results for the TNOs whose $grz$ photometry is presented here are tabulated in Table \ref{tableB}. Their orbital characteristics, apparent and absolute magnitudes are tabulated in Table \ref{tableA}.

Some of the key results for the surfaces of TNOs in the dynamical classes described in Section \ref{dclass} are:
\begin{itemize}[itemsep = 1pt, leftmargin=*]
\item The (21) TNOs dynamically classified as cold classicals in the $grz$ sample predominately populate $\mathcal{S}_{g \to r}$ of $\sim 18-31 $, and  $\mathcal{S}_{r \to z}$ of $\sim 9-27 $. There is one immediately apparent outlier, (612733) 2003 YU$_{179}$, with ($\mathcal{S}_{g \to r}$, $\mathcal{S}_{r \to z}$) = ($7.58 \pm 1.70, 13.61 \pm 2.45$). 2004 KE$_{19}$ is dynamically classified as cold classical which is also separated from the majority of the other cold classicals with ($\mathcal{S}_{g \to r}$, $\mathcal{S}_{r \to z}$) = ($18.80 \pm 1.30, 10.14 \pm 1.74$). The Col-OSSOS TNO 2013 UP$_{15}$ with ($\mathcal{S}_{g \to r}$, $\mathcal{S}_{r \to z}$) = ($22.895 \pm 1.181, 9.58 \pm 4.85$) lies outside the main cold classical grouping but has a large uncertainty in $\mathcal{S}_{r \to z}$. An additional observation at higher precision is needed to constrain the surface type of 2013 UP$_{15}$. The FaintIR:BrightIR surface class number ratio observed for the cold classicals is $\in$ [16:5, 19:2].

\item The (19) TNOs dynamically classified as non-resonant dynamically excited in the $grz$ sample are predominately found with $\mathcal{S}_{g \to r} \leq 20$. Five (of 19) of these TNOs extend into $\mathcal{S}_{g \to r} \geq 20$, where a number of them have spectral slopes consistent with those of the main cold classical grouping. (469333) 2000 PE$_{30}$ (2000 PE$_{30}$ hereafter) is dynamically classified as detached and shows a $\mathcal{S}_{g \to r}$ consistent with that of the Sun, and a $\mathcal{S}_{r \to z}$ just red of solar. The spectral slopes of 2000 PE$_{30}$ are consistent with that of 2013 UQ$_{15}$, a dynamically confirmed Haumea family member \citep{pike_haumea}. The dynamically classified hot classical (511552) 2014 UE$_{225}$ is consistent with having the reddest surface in the sample, with ($\mathcal{S}_{g \to r}$, $\mathcal{S}_{r \to z}$) = ($28.96 \pm 0.77, 26.40 \pm 2.08$) The FaintIR:BrightIR surface class number ratio observed for the non-resonant dynamically excited objects is $\in$ [5:14, 7:12].

\item The (14) plutinos show a wide range in $\mathcal{S}_{g \to r}$ $\mathcal{S}_{r \to z}$. These TNOs are predominately found with $\mathcal{S}_{g \to r} \leq 18$,  $\mathcal{S}_{r \to z} \leq 19$. 2007 HA$_{79}$ is consistent with having the reddest surface in the sample, with ($\mathcal{S}_{g \to r}$, $\mathcal{S}_{r \to z}$) = ($31.08 \pm 1.67,  30.255 \pm 3.452$). The FaintIR:BrightIR surface class number ratio observed for the plutinos is 3:11.

\item The $\mathcal{S}_{g \to r}$ \& $\mathcal{S}_{r \to z}$ spectral slopes of Mors and Somnus, the two components of the widely separated equal sized binary plutino Mors-Somnus, were measured individually and are reported in Table \ref{tableB}. Mors and Somnus show indistinguishable reflectance through $grz$. Therefore, Mors and Somnus show indistinguishable reflectance in wavelengths spanning $\sim 0.4$ to $5.3$ $\mu$m \citep{carol2024}. Mors and Somnus are both classified to have FaintIR surfaces.

\item The (7) TNOs in the $grz$ sample classified as resonant within the classical belt show a relatively uniform distribution in spectral slope space. These TNOs also show an even distribution in ($g-r$) \& ($r-J$) color-space as seen in \cite{PikeResonantSurfaces}. The FaintIR:BrightIR surface class number ratio observed for this dynamical class is $\in$ [2:5, 4:3].

\item The (4) TNOs in the $grz$ sample in the 4:3 resonance with Neptune span a wide range in $\mathcal{S}_{g \to r}$ \& $\mathcal{S}_{r \to z}$. 2013 US$_{15}$ had its ($g-r$) and ($r-z$) colors measured in Col-OSSOS, and is the only 4:3 with $\mathcal{S}_{g \to r}$ and $\mathcal{S}_{r \to z}$ spectral slopes that closely resemble those of the main cold classical grouping. (524457) 2002 FW$_6$ occupies a unique area in Figures \ref{figure:specslopes} \& \ref{figure:specslopes_bygroup} with ($\mathcal{S}_{g \to r}$ \& $\mathcal{S}_{r \to z}$) = ($18.37 \pm 1.84,  31.76 \pm 3.59$), although the uncertainties in these measurements are quite large. The FaintIR:BrightIR surface class number ratio observed for the 4:3s is $\in$ [1:3, 2:2].

\item The small sample of TNOs ($n = 3$) dynamically classified as in resonance with Neptune with semi-major axes larger than 49 au show spectral slopes in-between the main cold classical grouping and the more spectroscopically neutral TNOs. The FaintIR:BrightIR surface class number ratio observed for these bodies is 1:2.

\end{itemize}

\subsection{Comparing ($g-z$) Distributions by Dynamical Class}
\label{gMz}

\begin{figure}
\centering
 	\includegraphics[scale=0.6]{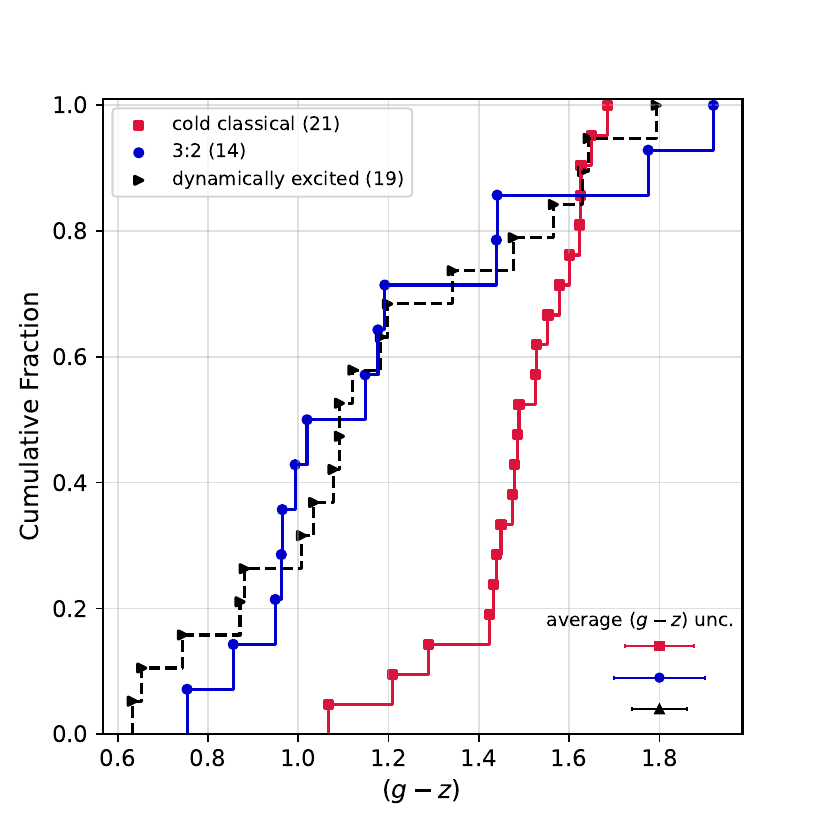}
	\caption{Cumulative distribution functions (CDFs)
 in ($g-z$) for the cold classical, plutino and non-resonant dynamically excited dynamical classes. The average uncertainty of ($g-z$) for each group is shown in the bottom right. Approximately 80\% of the plutinos, and 80\% of the non-resonant dynamically excited TNOs, have ($g-z$) $\lesssim$ 1.4 mags, while approximately 80\% of the cold classicals have ($g-z$) $\gtrsim$ 1.4 mags.
 \label{figure:PC1_gMz}
 }
\end{figure}

Significantly different distributions of $\mathcal{S}_{g \to r}$ and $\mathcal{S}_{r \to z}$ among the different dynamical classes implies that the TNOs in those dynamical classes have different distributions in their surface compositions. To directly compare these distributions, we use the Two-Sample Cramér-Von Mises statistic \citep[CVM,][]{2_samp_cvm} and the Two-Sample Anderson-Darling statistic \citep[AD,][]{2_samp_ad, k_samp_ad}. These two tests require one dimensional data. In Figures \ref{figure:specslopes} \& \ref{figure:specslopes_bygroup}, it is clear that the $\mathcal{S}_{g \to r}$ and $\mathcal{S}_{r \to z}$ of the TNOs in the sample have a tendency to cluster around the reddening line, which motivates use of the reddening line for the PC$^1$ \& PC$^2$ re-projection. There is a sparsely populated area that appears diagonal from the reddening line, which intersects the reddening line near ($\mathcal{S}_{g \to r}$, $\mathcal{S}_{r \to z}$) = ($17.5$, $17.5$). In Figure \ref{figure:specslopes_bygroup} (right), this sparsely populated area is linearized by a vertical line at PC$^1 \simeq 0.60$. In $grz$, we find that there is a strong linear correlation between PC$^1$ and ($g-z$). The PC$^1 \simeq 0.60$ separation seen in Figure \ref{figure:specslopes_bygroup} (right) becomes $(g-z) \simeq 1.35$ mags in Figure \ref{figure:PC1_gMz}. Motivated by the dispersion in ($g-z$), we use this color to compare the surface color distribution of the plutinos to that of the cold classicals and non-resonant dynamically hot TNOs. Figure \ref{figure:PC1_gMz} plots the ($g-z$) CDFs for the plutinos, cold classicals and non-resonant dynamically excited TNOs in the main $grz$ sample\footnote{The results of the statistical tests found below are consistent with the same analysis comparing PC$^1$ CDFs.}.

The CVM and AD tests make no assumptions on the profiles of the one dimensional distributions being compared. The Two-Sample Anderson-Darling rank statistic is a variation of the Two-Sample Cramér-Von Mises statistic in that it places a larger weight on the differences in the tails of the two CDFs. The AD statistic of interest is Equation 6 of \citet{k_samp_ad}, AD$_{kN}$ (AD hereafter), and the CVM statistic of interest is Equation 9 of \citet{2_samp_cvm}, T$_{2N}$ (T hereafter). The T-statistic was determined by pulling the value from the \texttt{stats.cramervonmises\_2samp()} function in the \texttt{SciPy} module. The significance of each statistic was evaluated using two techniques which we detail for the example of comparing the cold classicals to the plutinos in ($g-z$). 

The first significance technique pools the ($g-z$) values of the cold classicals and the plutinos into one group, uses the pooled values to randomly generate two new groups of the same sample sizes without replacement, and calculates both statistics using the two generated groups. This process was repeated 10000 times. The significance of each statistic can be determined by the number of runs out of 10000 that produced a statistic greater than the statistic measured between the true ($g-z$) colors of the cold classicals and the plutinos. The second technique utilizes the bootstrapping method first introduced in \citet{bootstrap}. We bootstrap re-sampled (re-sampled with replacement) the cold classical ($g-z$) color sample into a new sample of the size of the plutinos, and each statistic (T, AD) is calculated between the original cold classical ($g-z$) color sample and the bootstrap re-sampled cold classical ($g-z$) color sample. This process was repeated 10000 times. Each bootstrap re-sample of the cold classical ($g-z$) color values is thought to be another sample of the parent cold classical population's CDF in ($g-z$). Therefore, the values of both statistics calculated through this bootstrap technique represent the distribution or range of the AD and T statistics  when two samples share the same parent population CDF in ($g-z$). The significance of either statistic measured between the true ($g-z$) colors of the cold classicals and the plutinos can then be determined by the fraction of bootstrap runs that produced a statistic greater than the true statistic measured between the ($g-z$) colors of the cold classicals and the plutinos shown in Figure \ref{figure:PC1_gMz}.

The ($g-z$) distribution of the cold classicals ($n = 21$) were compared to that of the plutinos ($n = 14$). Since we added Mors-Somnus to the comparison target list because it was not from the surveys used to construct our sample and was known to have a cold classical-like surface in the optical \citep{ty430}, we did not include it in the list of plutinos for this analysis. The AD and T statistics measured by comparing the ($g-z$) distributions of the cold classicals and the plutinos were 5.47 and 1.14, respectively. The first significance technique by random grouping produced 0/10000 AD statistics, and  0/10000 T statistics as large as 5.47 and 1.14, respectively. The bootstrapping technique also produced 0/10000 AD statistics, and  0/10000 T statistics as large as 5.47 and 1.14, respectively. We therefore find that there is a $<0.01\%$ chance that the discrepancy seen between the ($g-z$) distributions of the cold classicals and the plutinos occurred by randomly sampling a shared ($g-z$) distribution into the sub-groups that we observed. This is evidence that at large, the planetesimals trapped into Neptune's 3:2 resonance were not sourced from the same parent planetesimal population of the cold classicals.

We next compare the ($g-z$) distributions of the plutinos ($n = 14$) and the non-resonant dynamically excited TNOs ($n = 19$) using the same technique.  The AD and T statistics measured by comparing the ($g-z$) distributions of the plutinos and the non-resonant dynamically excited TNOs were 0.33 and 0.05, respectively. The first significance technique by random grouping produced 5446/10000 ($54.46\%$) AD statistics, and 5860/10000 ($58.6\%$) T statistics as large as 0.33 and 0.05, respectively. The bootstrapping technique produced 4790/10000 ($47.9\%$) AD statistics, and 5826/10000 ($58.26 \%$) T statistics as large as 0.33 and 0.05, respectively. Therefore, we find that there is a $\gtrsim 50 \%$ chance that the ($g-z$) distributions of the non-resonant dynamically excited TNOs and the plutinos occurred by randomly sampling a shared ($g-z$) distribution into the sub-groups observed. These results are consistent with the hypothesis that the plutinos and the non-resonant dynamically excited objects share the same parent planetesimal population. We cannot rule out the possibility that the source region of these two groups was distinct while their surface color distributions were indistinguishable.

\subsection{FaintIR/BrightIR Surface Classifications}
\label{compclasses}

\begin{figure*}
\centering
 	\includegraphics[scale=0.6]{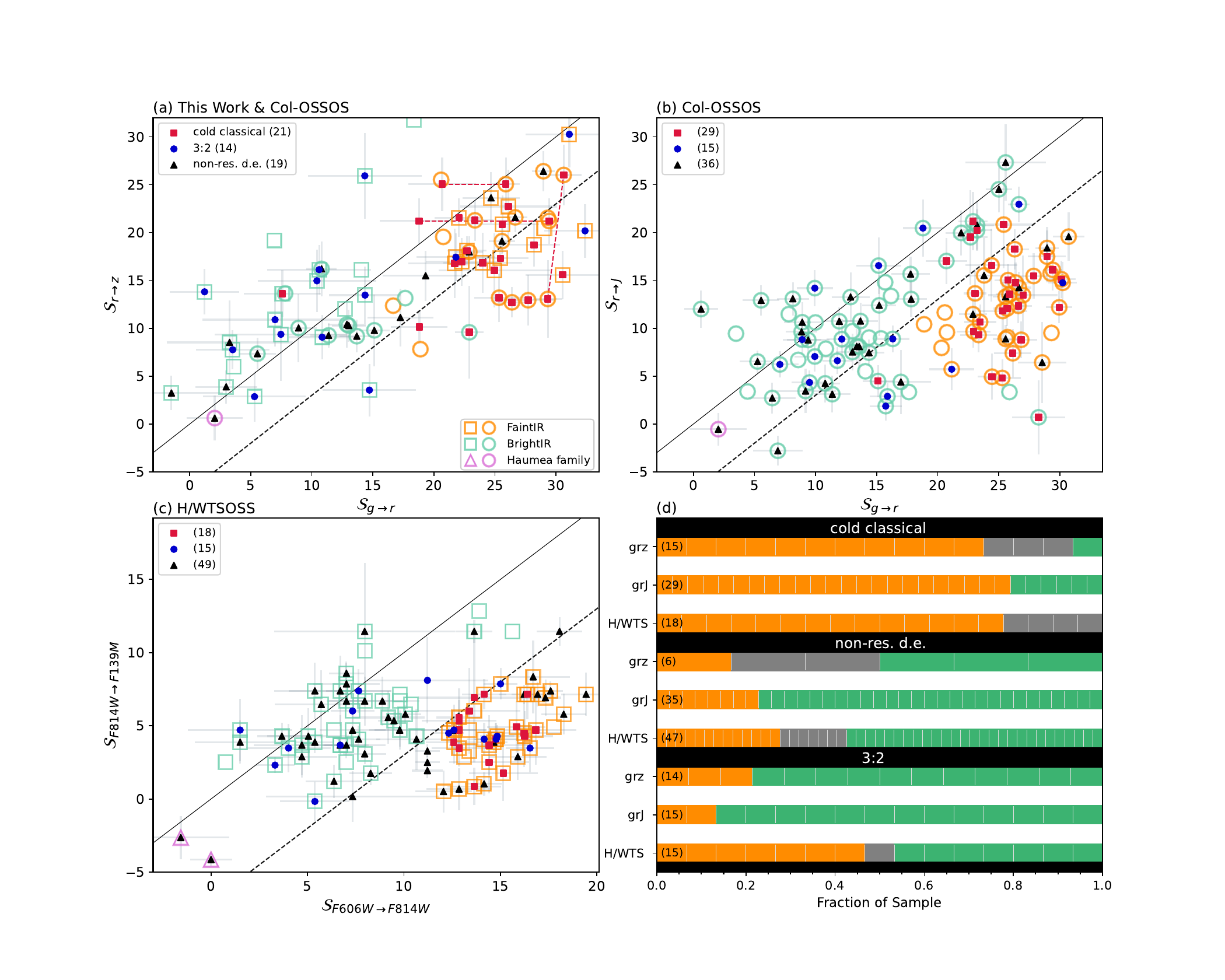}
	\caption{
 \textbf{(a)} Spectral slope space for the $grz$ dataset. In each panel, the solid black line indicates the reddening line and the dashed black line indicates a constant value of PC$^2 = -0.13$. Orange and blue-green outlines indicate FaintIR and BrightIR surface classifications respectively. Purple outline indicates that the TNO is a previously confirmed Haumea family member from \citet{pike_haumea}. Circle outline indicates surface classification from Col-OSSOS $grJ$ photometry \citep{Fraser2surfaces,PikeResonantSurfaces}. Square outlines indicate surface classification described in Section \ref{compclasses}. Points without surface classification outlines have an ambiguous surface class. TNOs with FaintIR/BrightIR classifications in other dynamical classes are plotted with with their surface classification only to avoid cluttering the figure. \textbf{(b)} Same as (a) but for the Col-OSSOS dataset. \textbf{(c)} Same as (a) but for the H/WTSOSS dataset. Triangle outline indicates the TNO is a previously confirmed Haumea family member from \citet{Schaller_haumea,Snodgrass_haumea}. \textbf{(d)} Apparent percentage of TNOs in each dynamical class in each program with surfaces classified as FaintIR and BrightIR using same color scheme as (a), but TNOs with ambiguous surface classifications are now indicated in grey. Haumea family members are not included. \label{figure:colorsinlit}
    }
\end{figure*}

In ($g-z$), the dynamically classified cold classicals stand out as having a unique surface color distribution in this wavelength range. However, we lose information when reducing the two dimensional ($g-r$) and ($r-z$) data to ($g-z$). A technique for distinguishing surface classes of TNOs was developed and refined in \citet{fraser2012} and \citet{Fraser2surfaces}, section 3.1.1 and 5 respectively. Two main TNO surface classes are hypothesized in \citet{Fraser2surfaces}, motivated by the bifurcated distributions of TNO colors in the Hubble/Wide Field Camera 3 F606W, F814W, and F139M filters \citep{fraser2012} and in $grJ$ \citep{Fraser2surfaces, PikeResonantSurfaces}. In the model, both compositional classes share a common optically neutral component, but have unique optically red components. The range of colors for a particular surface class are the result of different \% compositions of the two components and different transition wavelengths that define where the optical slope transitions to a near-IR slope \cite[see][Figure 7]{Fraser2surfaces}.  In \citet{Fraser2surfaces}, the FaintIR surface class is identified below PC$^2_{grJ} = -0.13$ and below PC$^1_{grJ} = 0.4$ (see Section \ref{measuringcolors} for description of PC$^1$ and PC$^2$), while the BrightIR surface class is identified outside of this region. The overwhelming majority of the dynamically classified cold classicals belong to the FaintIR class in $grJ$. To summarize, the FaintIR/BrightIR surface classification is more physically meaningful than a one dimensional ($g-z$) color.

Cataloging TNOs with FaintIR surfaces in other dynamical classes than the cold classicals can provide further insight into the origin and dynamical history of these planetesimals. Classifying the surfaces of the TNOs in our $grz$ sample as FaintIR/BrightIR first requires a well-motivated classification scheme for this filter set. In \cite[][Figure 7, bottom panel]{Fraser2surfaces}, the synthetic spectra of the model ($g-r$) \& ($r-J$) colors of the FaintIR and BrightIR classes were convolved with the Sloan $z$-band filter to produce ($r-z$) colors. The modeled BrightIR surface class dominates at the more neutral ($g-r$) \& ($r-z$) colors and hugs the reddening line at redder optical colors. The modeled FaintIR surface compositional class is present below the reddening line predominately at $(g-r)>0.75$ mags \cite[][Figure 7, bottom panel]{Fraser2surfaces}. In Figure \ref{figure:colorsinlit} (a) of the $grz$ sample, the Col-OSSOS TNOs with FaintIR \& BrightIR classifications are marked by orange and blue-green rings, respectively. Only cold classicals, non-resonant dynamically excited TNOs and plutinos are represented by symbols with uncertainties to avoid cluttering the figure. TNOs not in these dynamical classes are marked only by their surface classifications. The Col-OSSOS ($g-r$) \& ($r-J$) colors were converted to their corresponding spectral slopes and are shown in Figure \ref{figure:colorsinlit} (b)\footnote{We adopt the surface classifications of the Col-OSSOS TNOs as published in \cite[][Table 1]{PikeResonantSurfaces}.}. The Hubble Wide Field Camera 3 Test of Surfaces in the Outer Solar System (H/WTSOSS) dataset originally published in \citet{fraser2012} was also converted to spectral slope space and is shown in Figure \ref{figure:colorsinlit} (c). In these panels, the PC$^2 = -0.13$ lines  were converted into spectral slope space and are shown by dashed lines offset from the reddening line in their respective filter sets. As noted in \citet{Fraser2surfaces}, the majority of dynamically classified cold classical TNOs fall below the PC$^2 = -0.13$ in this H/WTSOSS filter set, similar to what was found in the Col-OSSOS $grJ$ colorspace \citep{Fraser2surfaces}. However, the same cannot be said for the $grz$ colorspace shown in Figure \ref{figure:colorsinlit} (a); many Col-OSSOS TNOs with FaintIR classified surfaces fall above and below the PC$^2 = -0.13$ line, and the cold classical TNOs are not primarily found below this line (though they almost all are below the reddening line with PC$^2 < 0$). Evidently, the PC$^2 = -0.13$ does not cleanly isolate the cold classical surface type in $grz$.

The directly measured ($g-r$) \& ($r-z$) colors of the (26) Col-OSSOS TNOs and the synthetic ($g-r$) \& ($r-z$) colors of the FaintIR \& BrightIR surface classes \citep[][Figure 7, bottom panel]{Fraser2surfaces} were used to guide a new classification scheme in $grz$. We classified TNO surfaces as FaintIR if their PC$^1_{grz}>0.55$ and PC$^2_{grz}<0$ (see Figure \ref{figure:specslopes}, right). The non-resonant dynamically excited TNO in the PC$^1_{grz} \sim 0.55-0.6$ region has an ambiguous classification. Similarly, the TNOs in Figure \ref{figure:colorsinlit} (a) in the $\mathcal{S}_{g \to r} \sim 15-20$ region straddling the PC$^2_{grz}=-0.13$ line have ambiguous surface classifications as FaintIR \& Bright IR Col-OSSOS TNOs are found on all sides. This classification scheme in $grz$ outlines in $\mathcal{S}_{g \to r}$ \& $\mathcal{S}_{r \to z}$ the areas occupied by Col-OSSOS FaintIR TNOs and the overwhelming majority of the cold classicals in this sample. The extent of the BrightIR surface class into the more $\mathcal{S}_{g \to r}$ red portion of Figure \ref{figure:colorsinlit} (a) is poorly understood. BrightIR TNOs are also found at the red end of $\mathcal{S}_{r \to J}$ at the red end of $\mathcal{S}_{g \to r}$ (Figure \ref{figure:colorsinlit} (b) and \citet{Fraser2surfaces}). Perhaps in $\mathcal{S}_{g \to r}$ \& $\mathcal{S}_{r \to z}$ there is a slight overlap of BrightIR and FaintIR surfaced TNOs along the reddening line at the red end of $\mathcal{S}_{g \to r}$, although there appear to be only a small portion of TNOs with optically red BrightIR surfaces \citep{schwamb2019,Fraser2surfaces}. Observations of these particular BrightIR TNOs are needed to better constrain the range of this surface class in $\mathcal{S}_{g \to r}$ \& $\mathcal{S}_{r \to z}$. 

Classifying the TNOs in the H/WTSOSS dataset into FaintIR/BrightIR surface types is motivated by the apparent bifurcation in $\mathcal{S}_{F606W \to F814W}$ and PC$^2 =-0.13$ as seen in \cite[][Figure 6]{Fraser2surfaces} and the two component Gaussian mixture model results published in \cite{Marsset23}. Objects with uncertainties that span the two groupings in this spectral slopes space had surfaces classified as uncertain. 

In Figure \ref{figure:colorsinlit} (d), the apparent percentage of FaintIR and BrightIR TNOs in each dynamical class in each program is shown. The unique filter set of the H/WTSOSS dataset and the overlap of BrightIR and FaintIR classes along the reddening line at the red end of $\mathcal{S}_{g \to r}$ in Figure \ref{figure:colorsinlit} (a) demands justification that our surface classifications of the H/WTSOSS dataset are compatible with that of the $grz$ and $grJ$ datasets. The $\mathcal{S}_{F606W \to F814W}$ of a TNO can be roughly described as a linear approximation of its spectrum through the $griz$ wavelength range weighted towards the pivot wavelengths of the filters. It resembles an axis roughly in the direction of the reddening line in $\mathcal{S}_{g \to r}$ \& $\mathcal{S}_{r \to z}$. The overlapping BrightIR and FaintIR surfaces at the red end of $\mathcal{S}_{g \to r}$ roughly translates to overlapping $\mathcal{S}_{F606W \to F814W}$. The F139M filter samples TNO reflectance further into the near-IR than $J$-band. Should the theoretical model FaintIR \& BrightIR surfaces reflect differently in the F814W $\to$ F139M wavelength range, we would expect to see differences in their $\mathcal{S}_{F814W \to F139M}$ at the red end of $\mathcal{S}_{F606W \to F814W}$. Here, we are most concerned with the accurate classification of the plutinos (blue circles in Figure \ref{figure:colorsinlit}). The majority of the H/WTSOSS plutinos classified as FaintIR fall within a small $\mathcal{S}_{F814W \to F139M}$ \& $\mathcal{S}_{F606W \to F814W}$ region dominated by a significant number of cold classicals, with the exception of one plutino showing a comparitively larger relative reflectance in the $F139M$ wavelength region.

\subsection{The $H_V \gtrsim 5$ Plutinos}
\label{main result}

A measurement of the FaintIR:BrightIR surface class ratio present in the plutino population can serve as an observational constraint for models of Neptune's migration and the pre-migration planetesimal disk structure. The observed FaintIR:BrightIR ratio from the $grz$ plutino sample is (3:11), similar to (2:13) measured for the Col-OSSOS plutinos. The H/WTSOSS plutinos exhibit a much larger FaintIR:BrightIR ratio, being $\in$ [7:8, 8:7]. 

\begin{figure}
\centering
 	\includegraphics[scale=0.65]{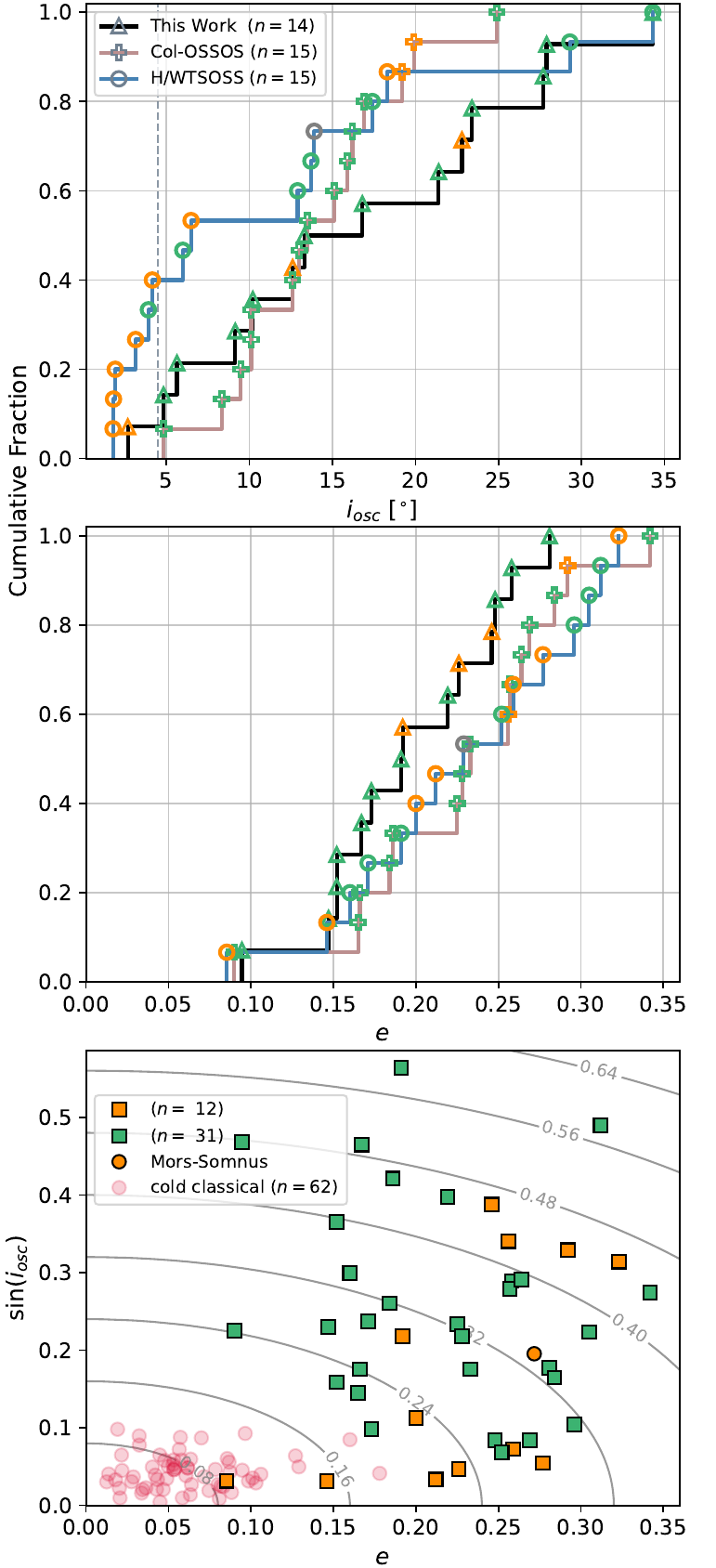}
	\caption{
 \textbf{Top/Middle:} CDFs in $i_{osc}$ and $e$ for the plutinos in the  $grz$, Col-OSSOS and H/WTSOSS datasets. \textbf{Bottom:} $\text{sin}(i_{osc})$ versus $e$ for the plutinos and cold classicals in the three datasets. The contours are iso-excitation lines, where $\epsilon = \sqrt{\text{sin}^2i_{osc} + e^2}$. Dynamical excitation increases radially from the lower left corner. The colors of the points represent surface classification, consistent with the color scheme in Figure \ref{figure:colorsinlit}. \label{figure:params}
 }
\end{figure}

In searching for a potential explanation for the large FaintIR fraction measured in the H/WTSOSS plutino sample, we investigated the different physical and orbital characteristics of the plutinos in each program. In Figure \ref{figure:params} (top/middle) we plot CDFs for the plutinos in each survey in $i_{osc}$ and $e$. Their orbital excitations $\epsilon$, where $\epsilon = \sqrt{\text{sin}^2i_{osc} + e^2}$, are shown in the bottom panel. We also compared cumulative distributions in $H_V$ across the datasets and found that they were virtually equivalent within the uncertainties.

Notably, (5/6) of the H/WTSOSS plutinos with $i_{osc} < 4.5^{\circ}$ are classified as FaintIR. The only other observed plutino from these programs with $i_{osc} < 4.5^{\circ}$ is in the $grz$ sample and its surface class is also FaintIR. While only a small sample of $i_{osc} < 4.5^{\circ}$ plutinos are considered in this analysis ($n=7$), the tendency for these low inclination plutinos to belong to the FaintIR class is evident (collectively, (6/7) plutinos with $i_{osc} < 4.5^{\circ}$ are classified as FaintIR). On the other hand, there are 36 plutinos across the three datasets with $i_{osc} > 4.5^{\circ}$ (2005 TV$_{189}$ is in both the $grz$ and H/WTSOSS datasets). The observed FaintIR:BrightIR ratio for $i_{osc} > 4.5^{\circ}$ plutinos across the three datasets are 2:11, 2:13 and $\in$ [2:7, 3:6] for $grz$, Col-OSSOS $grJ$ and H/WTSOSS respectively. 

We then compared the $i_{osc}$, $e$, and $\epsilon$ distributions of the plutinos in the FaintIR class to those in the BrightIR class. Across the $grz$, Col-OSSOS $grJ$ and H/WTSOSS datasets, there are (43) plutinos; (12) FaintIR, (30) BrightIR and (1) whose surface class is ambiguous. We make three assumptions: (\textbf{\textit{a}}) the FaintIR ($i_{osc}$, $e$, $\epsilon$) distribution are random samples of the intrinsic plutino FaintIR ($i_{osc}$, $e$, $\epsilon$) distributions (\textbf{\textit{b}}) the BrightIR ($i_{osc}$, $e$, $\epsilon$) distributions are random samples of the intrinsic plutino BrightIR ($i_{osc}$, $e$, $\epsilon$) distributions (\textbf{\textit{c}}) the FaintIR:BrightIR number ratio in our sample ($\in$ [12:31, 13:30]) is an accurate representation of this number ratio in the intrinsic plutino population. We compared the ($i_{osc}$, $e$, $\epsilon$) distribution of the plutinos in the FaintIR class to those in the BrightIR class following the statistical methodology described in Section \ref{gMz}. 

Under the above assumptions, the AD and CVM tests suggest the FaintIR and BrightIR $i_{osc}$ distributions are distinguishable at 99.3\% significance. In $\epsilon$, the AD and CVM tests suggest the FaintIR and BrightIR distributions are distinguishable at 97\% and 96\% significance respectively. The observed FaintIR plutino $i_{osc}$ and $\epsilon$ distributions extend down to lower values than the BrightIR plutinos, so the tendency for the higher significance in the AD test is due to the fact that it is more sensitive to differences in the tails of the distributions than the CVM test. Finally, the AD and CVM tests suggest the FaintIR and BrightIR $e$ distributions are indistinguishable. 

\section{Discussion and Conclusions}
\label{conclusions}

Across the $grz$, Col-OSSOS $grJ$ and H/WTSOSS datasets, there are (43) plutinos; (12) FaintIR, (30) BrightIR and (1) whose surface class is ambiguous, yielding a FaintIR:BrightIR number ratio $\in$ [12:31, 13:30]. The FaintIR:BrightIR ratio considering the three datasets more than doubles the (2:13) value measured in \cite{PikeResonantSurfaces}. Our analysis in Section \ref{main result} suggests that the larger FaintIR:BrightIR ratio considering the three datasets is the result of the $grz$ and H/WTSOSS plutino samples extending down to lower $i_{osc}$ than the Col-OSSOS plutinos alone. For the H/WTSOSS and $grz$ datasets, (6/7) of the plutinos with $i_{osc} < 4.5^{\circ}$ were FaintIR, while there are no $i_{osc} < 4.5^{\circ}$ plutinos in the Col-OSSOS $grJ$ sample. Under the assumption that the (43) plutinos are truly a random sample of the total 3:2 population, we found that the FaintIR and BrightIR $i_{osc}$ distributions are distinguishable at 99.3\% significance, but their distributions in $e$ are indistinguishable. In summary, plutinos with the FaintIR surface class have a different $i_{osc}$ distribution than the plutinos with the BrightIR surface class, where (6/7) of the plutinos in the H/WTSOSS and $grz$ datasets with $i_{osc} < 4.5^{\circ}$ are found to the have the FaintIR surface type.

Our finding that the FaintIR and BrightIR plutinos have different $i_{osc}$ distributions is consistent with the results published in \cite{Bernardinelli2025}. In that work, they use a TNO surface-color classification that is consistent with the FaintIR/BrightIR classifications used in this work, where the cold classicals are dominated by the `near-IR faint' type. Specifically, using a sample size of (49) plutinos they find strong evidence for inclination segregation of the near-IR faint and near-IR bright plutinos, where the near-IR faint plutinos are concentrated at lower inclinations.

The different $i_{osc}$ distributions of the FaintIR and BrightIR plutinos is evidence that the FaintIR plutinos are sourced from a different parent population than the BrightIR plutinos. For the plutino population as a whole, their total $i_{osc}$ distribution has been modeled by the probability profile $P(i)\propto \text{sin}(i)\text{exp}[-\frac{i^2}{2\sigma^2}]$ with an inclination width ($\sigma$) being $\sim$14° and an average inclination of $\sim$17.4° \cite[see, e.g.,][and references therein]{alexandersen2016}. \citet{gomes2000} and \cite{Balaji2023} have shown that the $i_{osc}$ distribution of the plutinos cannot be acquired through stability sculpting of planetesimals on initially cold orbits surrounding the current location of the 3:2 resonance ($\sim39.4$ au). Another excitation mechanism is responsible for the high $i_{osc}$ plutinos. 

Our findings then indicate that the FaintIR and BrightIR plutinos had, at large, different dynamical histories that resulted in their different $i_{osc}$ profiles. This can occur if the FaintIR and BrightIR planetesimals that were captured into the 3:2 resonance were radially partitioned in the primordial planetesimal disk \cite[e.g.][]{nesvorny2020, buchanan2022}. To be clear, there is already a relationship between surface type and inclination in the main belt between $\sim$42.5 au and the 2:1 resonance ($\sim$47.7 au); the current cold classical belt at $i_{free} \lesssim 5$° (roughly similar to $i_{osc}<4.5$°; free inclinations of resonant TNOs are not physically meaningful \citep{huang2022}) is dominated by the FaintIR surface type. The lowest inclination plutinos, also dominated by the FaintIR surface type, may be primordial cold classicals that formed at heliocentric distances $\lesssim 39.4$ au that were captured into the 3:2 resonance. In other words, the FaintIR plutinos with $i_{osc} < 4.5$° may be evidence that the primordial cold classical planetesimal disk was extended at least an additional $\sim$3 au in the past in the sunward direction from its current edge at $\sim42$ au. 

An alternative explanation for the FaintIR contribution in the observed plutinos cannot be ruled out. In the first explanation, the FaintIR plutinos are captured nearby 39.4 au (where they formed) and their low inclinations are sustained over $\sim$4.5 Gyr. However, preservation of initially low inclination orbits during/after capture into the 3:2 resonance is not guaranteed; the presence of the $\nu_{18}$, $\nu_8$, and Kozai secular resonances within/around the 3:2 act to raise inclinations \cite[e.g.][]{nesvorny2015b, volkmalhotra2019} and eccentricities of planetesimals before and/or during capture into the 3:2 resonance \citep{morbidelli1995, duncan95, morbidelli1997}. In the alternative scenario, the FaintIR plutinos are captured planetesimals that were scattered inwards from the cold classical population between $\sim42.5 - 48$ au (dominated by FaintIR surfaces), either by scattering capture during Neptune's migration \citep[e.g.][]{Kaib2016} or by a recent scattering event and temporary resonance sticking \citep[e.g.][]{lykawkamukai07}. The current centaur population with semi-major axes $\lesssim 30$ au, thought to be sourced from the scattering TNOs \citep{Malhotra2019}, exhibit a similar color-inclination relationship. Centaurs with cold classical-like surfaces have a colder inclination distribution in comparison to their less-red counterparts, while their eccentricity distributions are indistinguishable \cite[e.g.][]{tegler2016, marsset2019}. Since these centaurs have short stability timescales (a few 10s of Myr), the low inclination centaurs with cold classical-like surfaces must have a constant source to replenish this population, potentially being inward scattering cold classicals.  

Either mechanism for populating the plutinos with FaintIR surfaces must be consistent with our observations of the preponderance of FaintIR plutinos with $i_{osc}<4.5$°, where orbital excitation of these particular plutinos was primarily partitioned into eccentricity (see Figure \ref{figure:params}, bottom panel). Future simulations of FaintIR planetesimal capture into the 3:2 resonance could differentiate between the proposed mechanisms or conclude that both processes have occurred. The Legacy Survey of Space and Time (LSST) will discover additional plutinos on currently cold orbits \citep{Schwamb2023}, and constraints of their surface types would provide better statistics on the FaintIR:BrightIR ratio for plutinos with $i_{osc} < 4.5$°. In addition to the plutinos and the centaur populations, observations of TNOs in the 4:3 ($\sim$36.5 au) and 5:4 ($\sim$34.9 au) MMRs are significantly under-sampled today and would provide further observational constraints of the distribution of FaintIR surfaces in inner transneptunian space. 

Recent JWST NIRspec observations of TNO surfaces published in \citet{Pinilla-Alonso2025} point to three main TNO surface types. Although their cold classical sample is small, these TNOs appear to be dominated by the `cliff' spectral type whose unique spectral signature is attributed to complex organics and methanol ice. It is therefore likely that the FaintIR surface type, characteristic of the cold classicals in the present study, represents the spectral signature of the cliff spectral type at wavelengths $\sim 0.4$ to $0.9$ $\mu$m. Visible spectroscopy or spectrophotometry at shorter wavelengths is complimentary to these JWST observations. Future work bridging the gap between optical near-IR observations of TNOs and the JWST spectral types is critical to understanding the broader sample of TNOs and their evolutionary history.

\begin{acknowledgements}
The authors thank Wes Fraser, Estela Fernández-Valenzuela and Ben Proudfoot for helpful discussions that improved this manuscript.

The authors wish to acknowledge the land on which they live and carry out their research: Center for Astrophysics | Harvard \& Smithsonian is located on the traditional and ancestral land of the Massachusett, the original inhabitants of what is now known as Boston and Cambridge. We pay respect to the people of the Massachusett Tribe, past and present, and honor the land itself which remains sacred to the Massachusett People.

The authors acknowledge the sacred nature of Maunakea and appreciate the opportunity to observe from the mountain. CFHT is operated by the National Research Council (NRC) of Canada, the Institute National des Sciences de l’Universe of the Centre National de la Recherche Scientiﬁque (CNRS) of France, and the University of Hawaii, with L$i$DO receiving additional access due to contributions from the Institute of Astronomy and Astrophysics, Academia Sinica, Taiwan. This paper includes data gathered with the 6.5 meter Magellan Telescopes located at Las Campanas Observatory, Chile.

This work made use of the Pan-STARRS1 Surveys (PS1) and the PS1 public science archive which have been made possible through contributions by the Institute for Astronomy, the University of Hawaii, the Pan-STARRS Project Office, the Max-Planck Society and its participating institutes, the Max Planck Institute for Astronomy, Heidelberg and the Max Planck Institute for Extraterrestrial Physics, Garching, The Johns Hopkins University, Durham University, the University of Edinburgh, the Queen's University Belfast, the Harvard-Smithsonian Center for Astrophysics, the Las Cumbres Observatory Global Telescope Network Incorporated, the National Central University of Taiwan, the Space Telescope Science Institute, the National Aeronautics and Space Administration under Grant No. NNX08AR22G issued through the Planetary Science Division of the NASA Science Mission Directorate, the National Science Foundation Grant No. AST–1238877, the University of Maryland, Eotvos Lorand University (ELTE), the Los Alamos National Laboratory, and the Gordon and Betty Moore Foundation.

CC, REP, and MA acknowledge NASA Solar System Observations grant 80NSSC21K0289.
CC was supported in part by a Massachusetts Space Grant Consortium (MASGC) award and multiple travel grants from the University of Central Florida.
REP acknowledges NASA Emerging Worlds grant 80NSSC21K0376 and NASA Solar System Observations grant 80NSSC23K0680. SML was supported in part by NSERC Discovery Grant RGPIN-2020-04111.
\end{acknowledgements}

\facilities{Magellan Baade (IMACS), \\CFHT (MegaPrime/MegaCam)} 
\software{TRIPPy, Source Extractor, SciPy}

\clearpage
\appendix
\section{Color Terms}
\label{color terms}

\begin{figure*}[h!]
\centering
 	\includegraphics[scale=0.6]{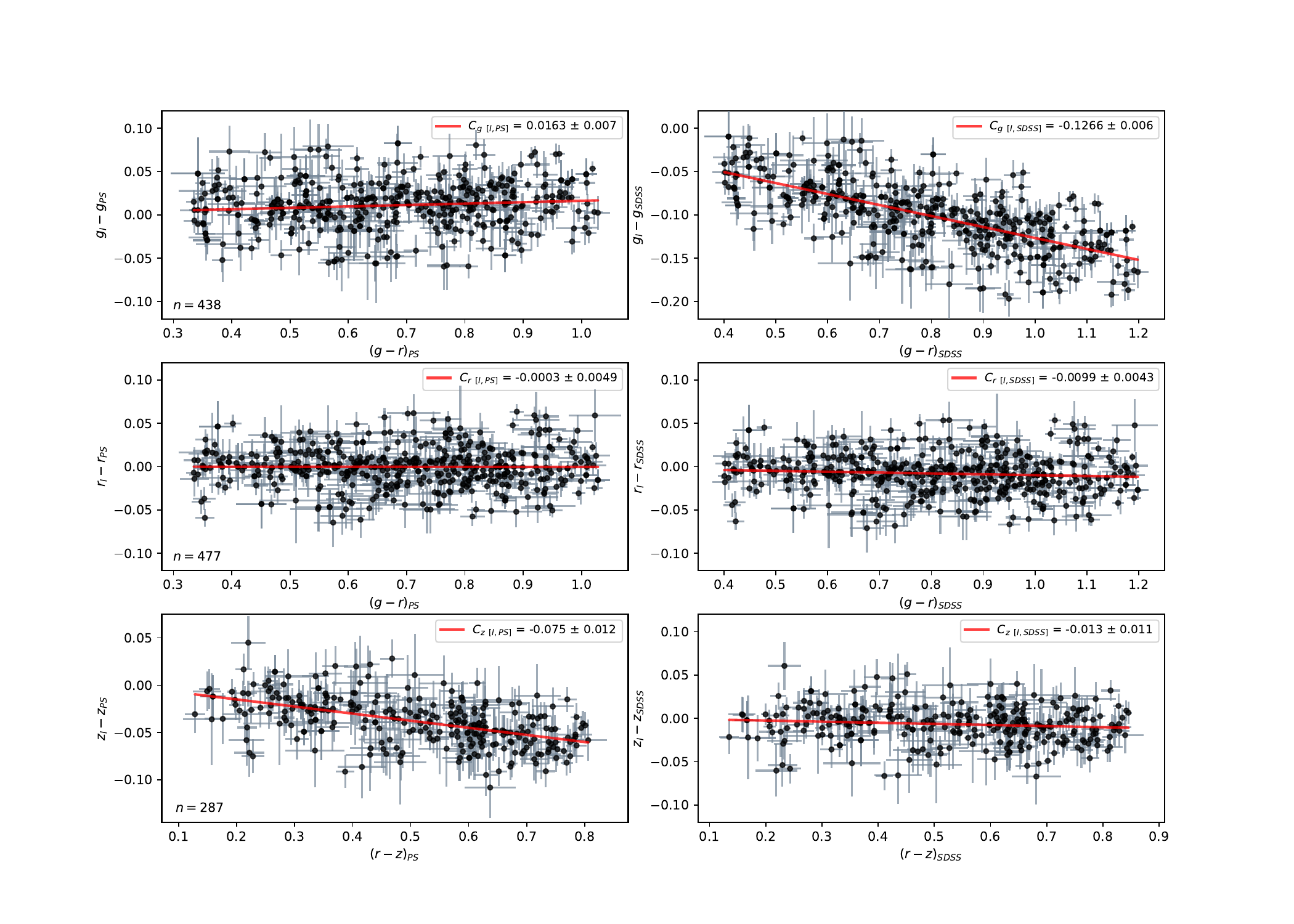}
	\caption{\label{figure:colorterms}
 Stellar magnitude and color data used to measure linear relations between IMACS and PS1 filters (left column) and IMACS and SDSS filters. The number of stars used to constrain the linear fits is noted in the bottom left of each panel in the left column (the same stars for a given filter type were used for IMACS$\to$PS1 and IMACS$\to$SDSS). See text for further details.  
 }
\end{figure*}

Using in-frame stars, our $grz$ aperture photometry was calibrated to the Pan-STARRS 1 \citep{PS1} photometric system and transformed to IMACS instrumental magnitudes. In-frame calibration stars were chosen by human inspection to ensure removal of galaxies, stars with cosmic ray contamination, stars with overlapping PSFs, stars near chip imperfections, stars approaching the non-linearity limit of the pixels, and SNRs $\geq 100$. We further filtered our calibration stars to have $0.4 \leq (g-r)_{SDSS} \leq 1.2$ and $0.05 \leq (r-z)_{SDSS} \leq 0.85$ as these colors are representative of the range of observed TNO colors. This resulted in $438$, $477$ and $287$ calibration stars for $g$, $r$ and $z$ respectively. The filter transformation equations between IMACS and Pan-STARRS 1 systems were calculated using the technique described in \cite{schwamb2019}; for each filter type ($g$, $r$, or $z$), a linear color term was calculated by simultaneously fitting the zeropoints of each image and the color term ($C$) between the two systems. Using the Pan-STARRS 1 to SDSS filter transformation equations from \citet{TonryPS1}, we also calculated the filter transformation equations between the IMACS and SDSS filter systems. The resulting filter transformation equations are summarized in Equations \ref{eq:I_PS} \& \ref{eq:I_SDSS} and visualized in Figure \ref{figure:colorterms}.

\begin{equation}
\label{eq:I_PS}
    \begin{cases}
        g_I = g_{PS} + 0.0163(\pm0.0070)\times(g-r)_{PS}\\
        r_I = r_{PS} - 0.0003(\pm0.0049)\times(g-r)_{PS}\\
        z_I = z_{PS} - 0.075(\pm0.012)\times(r-z)_{PS}\\
    \end{cases}   
\end{equation}

\begin{equation}
\label{eq:I_SDSS}
    \begin{cases} 
        g_I = g_{SDSS} - 0.1266(\pm0.0060)\times(g-r)_{SDSS}\\
        r_I = r_{SDSS} - 0.0099(\pm0.0043)\times(g-r)_{SDSS}\\
        z_I = z_{SDSS} - 0.013(\pm0.011)\times(r-z)_{SDSS}\\
    \end{cases}       
\end{equation}

\clearpage
\section{Tables}
\label{tables}

\begin{deluxetable}{llllllllllll}
\tablehead{Survey & Discovery & MPC & Discovery & $m_\text{disc.}$ & $\sigma_{m, \text{disc.}}$ & $m_{r,\text{est.}}$ & $H_{r,\text{est.}}$ & $a$ & $e$ & $i_{osc}$ & Status\\ ID&Survey&ID&Filter&mag&mag&mag&mag&au&&deg&Y/N}
\startdata
L4h09PD & CFEPS& 47932 (2000 GN171) & $g$ & 21.26 & 0.12 & 20.66 & 6.07 & 39.12 & 0.28 & 10.83& N\\
L4v13 &  CFEPS&681280 (2004 VV130) & $g$ & 22.72 & 0.11 & 22.12 & 6.99 & 39.77 & 0.19 & 23.88& N \\
0A07U07 & L$i$DO & 556130 (2014 KV101) & $w_{gri}$ & 22.33 & 0.14 & 22.27 & 7.34 & 39.11 & 0.26 & 16.82& Y \\
L3w07 &  CFEPS&  612658 (2003 TH58) & $g$ & 22.90 & 0.06 & 22.30 & 6.70 & 39.57 & 0.10 & 27.87& Y  \\
L4v18 &  CFEPS&  (2004 VY130) & $g$ & 22.95 & 0.22 & 22.35 & 7.81 & 39.69 & 0.28 & 10.19& Y  \\
L4h10PD &  CFEPS&  612029 (1995 HM5) & $g$ & 22.98 & 0.10 & 22.38 & 7.41 & 39.29 & 0.25 & 4.82& Y  \\
L4h08 &  CFEPS&  (2004 HZ78) & $g$ & 23.00 & 0.03 & 22.40 & 6.94 & 39.37 & 0.15 & 13.33& Y  \\
L5i06PD &  CFEPS&  306792 (2001 KQ77) & $g$ & 23.13 & 0.06 & 22.53 & 6.88 & 39.40 & 0.16 & 15.65& N \\
L3w01 &  CFEPS&  (2005 TV189) & $g$ & 23.24 & 0.09 & 22.64 & 7.50 & 39.63 & 0.19 & 34.33& Y \\
L4k11 &  CFEPS&  (2004 KC19) & $g$ & 23.35 & 0.18 & 22.75 & 7.90 & 39.10 & 0.23 & 5.65& N \\
L4m02 &  CFEPS&  (2004 MS8) & $g$ & 23.41 & 0.18 & 22.81 & 8.34 & 39.43 & 0.30 & 12.26& N \\
L3s02 &  CFEPS&  (2003 SO317) & $g$ & 23.47 & 0.14 & 22.87 & 7.81 & 39.62 & 0.27 & 6.55& N \\
$\substack{\text{L3s05}\\\text{o3l13PD}}$& $\substack{\text{CFEPS}\\\text{OSSOS}}$&  612621 (2003 SR317) & $g$ & 23.50 & 0.15 & 22.90 & 7.38 & 39.69 & 0.16 & 8.33& N \\
L4v09 &  CFEPS&  (2004 VX130) & $g$ & 23.56 & 0.08 & 22.96 & 7.54 & 39.76 & 0.21 & 5.73& N \\
L3h14 &  CFEPS&  612547 (2003 HD57) & $r$ & 22.98 & 0.10 & 22.98 & 7.73 & 39.18 & 0.17 & 5.63& Y  \\
L4h07 &  CFEPS&  (2004 HA79) & $g$ & 23.70 & 0.08 & 23.10 & 7.21 & 39.18 & 0.25 & 22.76& Y  \\
L4j11 &  CFEPS&  (2004 HX78) & $g$ & 23.71 & 0.18 & 23.11 & 7.78 & 39.25 & 0.15 & 16.30& N \\
L3h01 &  CFEPS&  (2004 FW164) & $r$ & 23.12 & 0.11 & 23.12 & 7.81 & 39.27 & 0.15 & 9.14& Y   \\
L3h11 &  CFEPS&  (2003 HA57) & $r$ & 23.16 & 0.13 & 23.16 & 7.94 & 39.22 & 0.17 & 27.68& Y   \\
L4h06 &  CFEPS&  (2004 HY78) & $g$ & 23.83 & 0.12 & 23.23 & 8.16 & 39.07 & 0.19 & 12.60& Y   \\
L5c11 &  CFEPS&  (2005 CD81) & $g$ & 23.87 & 0.28 & 23.27 & 6.71 & 39.24 & 0.15 & 21.33& Y   \\
L3h19 &  CFEPS&  (2003 HF57) & $r$ & 23.30 & 0.14 & 23.30 & 8.13 & 39.14 & 0.19 & 1.42& N \\
t806E15 & L$i$DO&  --- & $w_{gri}$ & 23.43 & 0.05 & 23.37 & 7.98 & 39.52 & 0.21 & 20.55& N \\
L4v12 &  CFEPS&  (2004 VZ130) & $g$ & 23.99 & 0.10 & 23.39 & 8.75 & 39.90 & 0.29 & 11.56& N \\
L4k01 &  CFEPS&  (2004 KB19) & $g$ & 24.01 & 0.12 & 23.41 & 7.38 & 39.34 & 0.22 & 17.19& N \\
L4h15 &  CFEPS&  (2004 HB79) & $g$ & 24.05 & 0.05 & 23.45 & 8.38 & 39.04 & 0.22 & 2.67& Y \\
3I37C09 & L$i$DO&  --- & $w_{gri}$ & 23.63 & 0.08 & 23.57 & 8.73 & 39.45 & 0.34 & 29.90& N \\
3C05C09 & L$i$DO&  --- & $w_{gri}$ & 23.90 & 0.17 & 23.84 & 9.53 & 39.55 & 0.33 & 19.40& N \\
0A10T05 & L$i$DO&  --- & $w_{gri}$ & 23.93 & 0.17 & 23.87 & 9.06 & 39.44 & 0.22 & 23.40& Y \\
2112C06 & L$i$DO&  --- & $w_{gri}$ & 23.97 & 0.28 & 23.91 & 8.73 & 39.46 & 0.23 & 30.31& N \\
\tablecaption{Target plutino sample description. \label{ttable}}
\enddata
\caption*{\footnotesize{\textbf{Note.}
$m_\text{disc.}$ is the apparent discovery magnitude, measured in the filter noted in the column to the left. $m_{r,\text{est.}}$ is the estimated apparent discovery magnitude transformed to the SDSS $r$ filter (see Section \ref{sample} for further details). $H_{r,\text{est.}}$ is the estimated absolute magnitude in SDSS $r$. The apparent discovery magnitude was first used to calculate a filter-dependent absolute magnitude, and then was converted to SDSS $r$ in the same way in which $m_{r,\text{est.}}$ was calculated. See the following table for more accurate absolute magnitudes and measured apparent magnitudes in SDSS $r$. The status column notes which of the target plutinos have been observed to date with satisfactory precision. 
}}
\end{deluxetable}

\begin{deluxetable}{lllllllllll}
\tablehead{Survey & Discovery&  MPC & Dynamical &  $m_{r,\text{meas.}}$ &  $\sigma_{m_{r,\text{meas.}}}$ &  $H_{r,\text{sur.}}$ &     $a$ &    $e$ &     $i$ &  $i_{\text{free}}$ \\ ID&Survey&ID&Class&mag&mag&mag&au&&deg&deg}
\startdata
o5p042PD & OSSOS&524457 (2002 FW6) & 4:3 & 23.66 & 0.03 & 7.26 & 36.19 & 0.12 & 3.54 & --- \\
L4h14 & CFEPS&(2004 HM79) & 4:3 & 23.51 & 0.03 & 6.94 & 36.28 & 0.08 & 1.18 & --- \\
L7a10 & CFEPS&(2005 GH228) & 4:3 & 23.27 & 0.04 & 8.27 & 36.62 & 0.19 & 17.15 & --- \\
L4o01 & CFEPS&(2004 OP15) & HC & 22.61 & 0.02 & 6.59 & 38.62 & 0.06 & 22.98 & --- \\
L4h15 & CFEPS&(2004 HB79) & 3:2 & 24.03 & 0.02 & 8.20 & 39.04 & 0.22 & 2.67 & --- \\
L4h06 & CFEPS&(2004 HY78) & 3:2 & 23.66 & 0.02 & 7.62 & 39.07 & 0.19 & 12.60 & --- \\
0A07U07 & L$i$DO&556130 (2014 KV101) & 3:2 & 22.43 & 0.02 & 7.34 & 39.11 & 0.26 & 16.82 & --- \\
L3h14 & CFEPS&612547 (2003 HD57) & 3:2 & 23.62 & 0.05 & 7.73 & 39.18 & 0.17 & 5.63 & --- \\
L4h07 & CFEPS&(2004 HA79) & 3:2 & 23.65 & 0.03 & 6.71 & 39.18 & 0.25 & 22.76 & --- \\
L3h11 & CFEPS&(2003 HA57) & 3:2 & 23.48 & 0.03 & 7.94 & 39.22 & 0.17 & 27.68 & --- \\
L5c11 & CFEPS&(2005 CD81) & 3:2 & 23.30 & 0.02 & 6.61 & 39.24 & 0.15 & 21.33 & --- \\
L3h01 & CFEPS&(2004 FW164) & 3:2 & 23.43 & 0.02 & 7.81 & 39.27 & 0.15 & 9.14 & --- \\
L4h10PD & CFEPS&612029 (1995 HM5)& 3:2 & 22.61 & 0.02 & 7.51 & 39.29 & 0.25 & 4.82 & --- \\
L4h08 & CFEPS&(2004 HZ78) & 3:2 & 22.81 & 0.02 & 7.08 & 39.37 & 0.15 & 13.33 & --- \\
0A10T05 & L$i$DO&--- & 3:2 & 23.86 & 0.02 & 9.06 & 39.44 & 0.22 & 23.40 & --- \\
L3w07 & CFEPS&612658 (2003 TH58) & 3:2 & 22.93 & 0.03 & 6.77 & 39.57 & 0.10 & 27.87 & --- \\
1K03U11 & L$i$DO&--- & HC & 23.75 & 0.02 & 7.95 & 39.58 & 0.02 & 18.73 & --- \\
L3w01 & CFEPS&(2005 TV189) & 3:2 & 22.75 & 0.03 & 7.40 & 39.63 & 0.19 & 34.33 & --- \\
L4v18 & CFEPS&(2004 VY130) & 3:2 & 22.84 & 0.03 & 7.85 & 39.69 & 0.28 & 10.19 & --- \\
L5c08 & CFEPS&434709 (2006 CJ69) & 5:3 & 22.61 & 0.02 & 7.02 & 42.08 & 0.23 & 17.91 & --- \\
L5c13PD & CFEPS&612086 (1999 CX131) & 5:3 & 22.86 & 0.03 & 6.95 & 42.15 & 0.23 & 9.75 & --- \\
L3q02PD & CFEPS&385266 (2001 QB298) & CC & 22.90 & 0.02 & 6.36 & 42.81 & 0.10 & 1.80 & 3.80 \\
L4m03 & CFEPS&(2004 MT8) & CC & 23.34 & 0.03 & 6.30 & 43.10 & 0.04 & 2.24 & 0.45 \\
L5c07PD & CFEPS&308634 (2005 XU100)& HC & 22.75 & 0.02 & 5.94 & 43.42 & 0.11 & 7.86 & 9.00 \\
L3y05 & CFEPS&2003 YS179) & CC & 23.29 & 0.02 & 6.50 & 43.67 & 0.02 & 3.72 & 1.94 \\
L4n03 & CFEPS&(2004 OQ15) & 7:4 & 23.25 & 0.03 & 7.04 & 43.72 & 0.13 & 9.74 & --- \\
L3h13 & CFEPS&(2003 HH57) & CC & 23.54 & 0.04 & 6.93 & 43.76 & 0.08 & 1.44 & 1.27 \\
o3e22 & OSSOS &500835 (2013 GN137) & CC & 23.04 & 0.02 & 6.70 & 43.86 & 0.06 & 2.77 & 1.10 \\
L4p06PD & CFEPS&275809 (2001 QY297) & CC & 21.89 & 0.02 & 5.14 & 43.94 & 0.08 & 1.55 & 0.86 \\
L5c22 & CFEPS&(2007 DS101) & CC & 22.77 & 0.03 & 6.11 & 43.95 & 0.08 & 1.39 & 1.57 \\
L4k02 & CFEPS&(2004 KE19) & CC & 22.83 & 0.02 & 6.10 & 44.14 & 0.05 & 1.18 & 2.86 \\
L4v03 & CFEPS&609221 (2004 VC131) & CC & 22.22 & 0.02 & 5.78 & 44.31 & 0.08 & 0.49 & 1.44 \\
L3w02 & CFEPS&183595 (2003 TG58) & CC & 22.65 & 0.03 & 6.11 & 44.85 & 0.11 & 1.65 & 3.43 \\
o3e30PD & OSSOS&(2013 EM149) & CC & 23.14 & 0.02 & 6.59 & 45.01 & 0.05 & 2.64 & 2.78 \\
o5p098PDp & OSSOS&525462 (2005 EO304) & CC & 22.62 & 0.02 & 6.04 & 45.40 & 0.06 & 3.42 & 1.74 \\
o5c073 & OSSOS&(2015 VT168) & CC & 23.21 & 0.04 & 6.51 & 45.67 & 0.04 & 1.20 & 2.63 \\
L4j07 & CFEPS&(2004 HD79) & CC & 22.33 & 0.01 & 5.23 & 45.72 & 0.04 & 1.30 & 1.60 \\
L7a04PD & CFEPS&524435 (2002 CY248) & HC & 22.55 & 0.02 & 5.15 & 46.22 & 0.14 & 7.02 & 8.56 \\
L3y03 & CFEPS&612733 (2003 YU179) & CC & 23.25 & 0.03 & 6.90 & 46.87 & 0.16 & 4.85 & 3.54 \\
L3y09 & CFEPS&(2003 YV179) & HC & 23.04 & 0.01 & 6.54 & 47.09 & 0.22 & 15.55 & 13.97 \\
L4p04PD & CFEPS&469333 (2000 PE30) & D & 22.27 & 0.02 & 5.82 & 54.46 & 0.34 & 18.42 & --- \\
L3f04PD & CFEPS&60621 (2000 FE8) & 5:2 & 23.22 & 0.03 & 6.31 & 54.99 & 0.40 & 5.88 & --- \\
\midrule 
   & $\substack{\text{Sheppard \&}\\\text{Trujillo}^{(a)}}$ & 341520 Mors-Somnus (2007 TY430)          & & & & 6.16 & & &  & \\
--- &  &       Mors &    3:2 &  21.73 &    0.02 & --- & 39.60 & 0.27 & 11.30 &  --- \\
--- &  &     Somnus &    3:2 &  22.05 &    0.02 & --- & 39.60 & 0.27 & 11.30 &  --- \\
\tablecaption{Description of the TNOs observed in our Magellan Baade program.
\label{tableA}}
\enddata
\caption*{\footnotesize{\textbf{Note.} (a) \cite{sheppardtrujillo10a}. Dynamical Class Key: HC = Hot Classical; CC = Cold Classical; D = Detached; e.g. 4:3 denotes the TNO is in the 4:3 mean motion resonance (see Section \ref{dclass} for dynamical classification scheme). $m_{r,\text{meas.}}$ is the average SDSS $r$-band apparent magnitude measured during our observations. $H_{r,\text{sur.}}$ is the surmised SDSS $r$-band absolute magnitude. This absolute magnitude is derived from the discovery apparent magnitude, where filter transformations to SDSS $r$-band utilized the ($g-r$) color measured in this work. $i_{free}$ measurements were taken from \cite{huang2022} and \cite{vanLaerhoven2019}. 
}}
\end{deluxetable}

\begin{deluxetable}{llllllllllllll}
\tablehead{Survey &  MPC &   $g-r$ &  $\sigma_{g-r}$ &   $r-z$ &  $\sigma_{r-z}$ & $\mathcal{S}_{g \to r}$ & $\sigma_{\mathcal{S}_{g \to r}}$ & $\mathcal{S}_{r \to z}$ & $\sigma_{\mathcal{S}_{r \to z}}$ &   PC$^1$ &  $\sigma_{\text{PC}^1}$ &    PC$^2$ &  $\sigma_{\text{PC}^2}$\\ID&ID&mag&mag&mag&mag&&&&&&&&}
\startdata
o5p042PD & 524457 (2002 FW6) & 0.781 & 0.080 & 0.844 & 0.055 & 18.370 & 3.800 & 31.760 & 3.590 & 0.724 & 0.057 & 0.189 & 0.049 \\
L4h14 & (2004 HM79) & 0.780 & 0.094 & 0.439 & 0.065 & 18.330 & 4.440 & 11.040 & 2.890 & 0.457 & 0.069 & -0.109 & 0.070 \\
L7a10 & (2005 GH228) & 0.499 & 0.081 & 0.312 & 0.093 & 3.580 & 4.970 & 5.980 & 3.590 & 0.177 & 0.077 & 0.038 & 0.068 \\
L4o01 & (2004 OP15) & 0.489 & 0.037 & 0.255 & 0.045 & 2.980 & 2.530 & 3.880 & 1.750 & 0.122 & 0.038 & 0.017 & 0.032 \\
L4h06 & (2004 HY78) & 1.139 & 0.099 & 0.636 & 0.051 & 32.370 & 3.320 & 20.170 & 2.790 & 0.900 & 0.093 & -0.185 & 0.052 \\
L4h15 & (2004 HB79) & 0.858 & 0.073 & 0.581 & 0.048 & 21.790 & 3.240 & 17.420 & 2.520 & 0.612 & 0.057 & -0.075 & 0.048 \\
L4h07 & (2004 HA79) & 1.100 & 0.103 & 0.819 & 0.054 & 31.080 & 3.590 & 30.260 & 3.450 & 0.966 & 0.091 & -0.009 & 0.051 \\
L5c11 & (2005 CD81) & 0.697 & 0.056 & 0.495 & 0.064 & 14.360 & 2.940 & 13.460 & 2.960 & 0.442 & 0.048 & -0.010 & 0.043 \\
0A07U07 & 556130 (2014 KV101) & 0.628 & 0.047 & 0.391 & 0.053 & 10.840 & 2.710 & 9.070 & 2.290 & 0.315 & 0.043 & -0.023 & 0.038 \\
L3h11 & (2003 HA57) & 0.696 & 0.092 & 0.744 & 0.077 & 14.330 & 4.650 & 25.920 & 4.460 & 0.609 & 0.061 & 0.167 & 0.061 \\
L3h14 & 612547 (2003 HD57) & 0.566 & 0.095 & 0.399 & 0.118 & 7.460 & 5.430 & 9.360 & 4.860 & 0.284 & 0.093 & 0.031 & 0.077 \\
L3h01 & (2004 FW164) & 0.623 & 0.078 & 0.553 & 0.051 & 10.590 & 4.260 & 16.120 & 2.580 & 0.436 & 0.049 & 0.084 & 0.058 \\
L4h10PD & 612029 (1995 HM5) & 0.497 & 0.046 & 0.358 & 0.048 & 3.520 & 3.000 & 7.750 & 2.060 & 0.216 & 0.040 & 0.064 & 0.038 \\
L4h08 & (2004 HZ78) & 0.459 & 0.042 & 0.503 & 0.049 & 1.210 & 2.850 & 13.810 & 2.360 & 0.303 & 0.035 & 0.179 & 0.033 \\
0A10T05 & --- & 0.620 & 0.089 & 0.528 & 0.068 & 10.410 & 4.850 & 14.960 & 3.240 & 0.415 & 0.057 & 0.070 & 0.068 \\
L3w07 & 612658 (2003 TH58) & 0.528 & 0.070 & 0.226 & 0.072 & 5.300 & 4.210 & 2.860 & 2.600 & 0.119 & 0.063 & -0.032 & 0.060 \\
L3w01 & (2005 TV189) & 0.704 & 0.075 & 0.245 & 0.075 & 14.710 & 3.810 & 3.540 & 2.740 & 0.238 & 0.069 & -0.170 & 0.061 \\
1K03U11 & --- & 0.493 & 0.084 & 0.378 & 0.108 & 3.260 & 5.160 & 8.520 & 4.380 & 0.228 & 0.085 & 0.079 & 0.069 \\
L4v18 & (2004 VY130) & 0.557 & 0.104 & 0.436 & 0.063 & 6.990 & 5.970 & 10.890 & 2.770 & 0.309 & 0.061 & 0.060 & 0.083 \\
L5c08 & 434709 (2006 CJ69) & 1.042 & 0.056 & 0.641 & 0.039 & 29.030 & 2.140 & 20.410 & 2.230 & 0.814 & 0.049 & -0.129 & 0.032 \\
L5c13PD & 612086 (1999 CX131) & 0.691 & 0.067 & 0.553 & 0.065 & 14.060 & 3.510 & 16.110 & 3.200 & 0.478 & 0.048 & 0.034 & 0.049 \\
L3q02PD & 385266 (2001 QB298) & 1.085 & 0.085 & 0.542 & 0.051 & 30.540 & 3.040 & 15.570 & 2.530 & 0.789 & 0.081 & -0.235 & 0.049 \\
L4m03 & (2004 MT8) & 0.856 & 0.087 & 0.567 & 0.074 & 21.710 & 3.810 & 16.770 & 3.630 & 0.609 & 0.067 & -0.072 & 0.058 \\
L5c07PD & 308634 (2005 XU100) & 0.756 & 0.054 & 0.441 & 0.069 & 17.250 & 2.690 & 11.130 & 3.020 & 0.442 & 0.054 & -0.090 & 0.046 \\
L3y05 & (2003 YS179) & 0.869 & 0.066 & 0.571 & 0.049 & 22.260 & 2.930 & 16.950 & 2.500 & 0.621 & 0.053 & -0.078 & 0.044 \\
L3h13 & (2003 HH57) & 0.964 & 0.100 & 0.686 & 0.070 & 26.100 & 3.960 & 22.710 & 3.850 & 0.773 & 0.084 & -0.047 & 0.059 \\
o3e22 & 500835 (2013 GN137) & 0.789 & 0.068 & 0.657 & 0.042 & 18.770 & 3.220 & 21.200 & 2.380 & 0.618 & 0.049 & 0.039 & 0.045 \\
L4n03 & (2004 OQ15) & 0.556 & 0.064 & 0.617 & 0.062 & 6.920 & 3.790 & 19.180 & 3.240 & 0.439 & 0.042 & 0.176 & 0.048 \\
L5c22 & (2007 DS101) & 0.951 & 0.074 & 0.650 & 0.056 & 25.600 & 3.020 & 20.850 & 3.030 & 0.742 & 0.061 & -0.068 & 0.045 \\
L4p06PD & 275809 (2001 QY297) & 0.864 & 0.042 & 0.663 & 0.033 & 22.050 & 1.990 & 21.550 & 2.000 & 0.679 & 0.032 & -0.005 & 0.027 \\
L4k02 & (2004 KE19) & 0.790 & 0.047 & 0.418 & 0.037 & 18.800 & 2.320 & 10.140 & 1.740 & 0.445 & 0.034 & -0.131 & 0.035 \\
L4v03 & 609221 (2004 VC131) & 0.947 & 0.055 & 0.578 & 0.037 & 25.440 & 2.320 & 17.300 & 2.010 & 0.691 & 0.047 & -0.122 & 0.035 \\
L3w02 & 183595 (2003 TG58) & 0.910 & 0.077 & 0.568 & 0.063 & 23.990 & 3.240 & 16.830 & 3.160 & 0.654 & 0.062 & -0.107 & 0.050 \\
o3e30PD & (2013 EM149) & 0.832 & 0.060 & 0.729 & 0.047 & 20.670 & 2.760 & 25.050 & 2.800 & 0.697 & 0.046 & 0.067 & 0.037 \\
o5p098PDp & 525462 (2005 EO304) & 1.019 & 0.054 & 0.607 & 0.041 & 28.190 & 2.110 & 18.710 & 2.270 & 0.773 & 0.046 & -0.143 & 0.034 \\
o5c073 & (2015 VT168) & 0.934 & 0.113 & 0.552 & 0.081 & 24.940 & 4.570 & 16.050 & 3.890 & 0.663 & 0.095 & -0.135 & 0.072 \\
L4j07 & (2004 HD79) & 0.880 & 0.032 & 0.595 & 0.024 & 22.710 & 1.590 & 18.100 & 1.520 & 0.648 & 0.026 & -0.067 & 0.021 \\
L7a04PD & 524435 (2002 CY248) & 0.927 & 0.054 & 0.702 & 0.046 & 24.670 & 2.330 & 23.620 & 2.700 & 0.755 & 0.043 & -0.012 & 0.037 \\
L3y03 & 612733 (2003 YU179) & 0.568 & 0.053 & 0.499 & 0.051 & 7.580 & 3.160 & 13.610 & 2.450 & 0.361 & 0.038 & 0.091 & 0.041 \\
L3y09 & (2003 YV179) & 0.801 & 0.075 & 0.540 & 0.053 & 19.320 & 3.500 & 15.480 & 2.630 & 0.550 & 0.057 & -0.055 & 0.052 \\
L4p04PD & 469333 (2000 PE30) & 0.416 & 0.062 & 0.237 & 0.047 & -1.500 & 4.210 & 3.240 & 1.780 & 0.072 & 0.039 & 0.070 & 0.054 \\
L3f04PD & 60621 (2000 FE8) & 0.664 & 0.093 & 0.462 & 0.094 & 12.730 & 4.850 & 12.020 & 4.130 & 0.394 & 0.072 & -0.008 & 0.071 \\
\midrule
     &  341520 Mors-Somnus (2007 TY430) &      & & & & & & & & & & & \\
--- &         Mors & 0.915 & 0.042 & 0.544 & 0.047 & 24.17 &  1.89 & 15.68 &  2.37 & 0.639 & 0.035 & -0.128 & 0.036 \\
--- &       Somnus & 0.910 & 0.049 & 0.638 & 0.063 & 23.97 &  2.16 & 20.25 &  3.35 & 0.610 & 0.041 & -0.053 & 0.049 \\ 
\tablecaption{ Measured surface parameters for the TNOs in our Magellan Baade program. 
\label{tableB}}
\enddata
\caption*{\footnotesize{
\textbf{Note.} See Section \ref{measuringcolors} for description of spectral slope $\mathcal{S}$ and PC$^1$ PC$^2$. 
}}
\end{deluxetable}

\clearpage

\bibliography{main}{}
\end{document}